\documentclass[useAMS,usenatbib]{mn2e}
\usepackage{times}
\usepackage{psfig}
\usepackage{graphicx}
\usepackage[usenames,dvips]{color}

\newcommand{\Fewbody}{{\tt Fewbody}}

\title[The Evolution of Binary Fractions in Globular Clusters]{The Evolution of Binary Fractions in Globular Clusters}
\author[N Ivanova et al.]
{N.\ Ivanova $^1$\thanks{E-mail:nata@northwestern.edu} , 
K.\ Belczynski$^{2}$\thanks{Tombaugh Postdoctoral Fellow}, J.~M.\ Fregeau$^1$, \& F.~A.\ Rasio$^1$\\
$^1$Northwestern University, Dept of Physics \& Astronomy,  Evanston, IL 60208, USA\\
$^2$Present address: New Mexico State University, Department of Astronomy,  1320 Frenger Mall,
 Las Cruces, New Mexico 88003-8001, USA
}

\begin{document}

\maketitle

\label{firstpage}

\begin{abstract}{
We study the evolution of binary stars in globular
clusters using a new Monte Carlo approach combining a population  synthesis
code ({\tt StarTrack}), and
a simple treatment of dynamical interactions
in the dense cluster core using a new tool for computing 3-body
and 4-body interactions ({\tt Fewbody}).
We find that the combination of stellar evolution and dynamical  interactions
(binary--single and binary--binary) leads to a rapid depletion of the  binary
population in the cluster core.
The {\em maximum\/} binary fraction today in the core of a typical  dense cluster
like 47~Tuc, assuming an initial binary fraction of 100\%, is only  about 5--10\%.
We show that this is in good agreement with recent {\em HST\/}  observations of close
binaries in the core of 47~Tuc, provided that a realistic distribution  of
binary periods is used to interpret the results. Our findings also have
important consequences for the dynamical modeling of globular clusters,
suggesting that ``realistic models'' should incorporate much larger
initial binary fractions than has usually been done in the past.
}
\end{abstract}


\section{Introduction}

Binary stars play a fundamental role in the evolution of
globular clusters for at least two important reasons.
First, the evolution of stars in binaries, whether in a
cluster or in the galactic field, can be very different from
the evolution of the same stars in isolation.
In a dense environment like a globular cluster, this difference is exacerbated by
dynamical encounters, which affect binaries 
much more than single stars.
Second, binary stars crucially affect the dynamical evolution of
globular clusters, providing (through inelastic collisions) the source
of energy that supports them against gravothermal collapse
\citep{GoodmanHut_89,Gao_FP_91,2003ApJ...593..772F}.
In the ``binary burning'' phase, a cluster can remain in quasi-thermal  equilibrium
with nearly constant core density and velocity dispersion for many
relaxation times, in a similar way to that in which a star can maintain  itself
in thermal equilibrium for many Kelvin-Helmholtz times by burning
hydrogen in its core.
The binary fraction\footnote{Throughout this paper, the ``binary  fraction''
among a particular group of objects is defined as the number of objects  that
are binaries divided by the total number of objects. So, for example,
a primordial binary fraction of 50\% implies that 2/3 of main-sequence  stars
are in binaries initially. However, a binary fraction of 50\% among  white
dwarfs later on does not imply that 2/3 of white dwarfs are in  binaries, as
some white dwarf companions may be main-sequence stars.}
(and the initial, primordial binary fraction in particular),
is therefore one of the most important parameters that determine
the evolution of globular clusters.
However, most previous dynamical studies of globular clusters---even those
including binaries---have neglected stellar evolution,
which can significantly impact the properties and survival
of binaries and hence the reservoir of energy they provide.

At present, there are very few direct measurements of binary fractions
in clusters.  However, even early observations showed that binary
fractions in globular cluster cores are smaller than in the solar
neighborhood \citep[e.g.,][]{Cote_M22_96}.  Recent Hubble Space
Telescope (HST) observations have provided further constraints
on the binary fractions in many globular clusters
\citep{2002AJ....123.2541B,1997ApJ...474..701R}.  The measured binary
fractions in dense cluster cores are found to be {\em very small\/}.
As an example, the upper limit on the core binary fraction of NGC 6397
is only 5-7\% \citep{CoolBolton_NGC6397_02}.
On the other hand, in very sparse clusters, like NGC 288  \citep{2002AJ....123.1509B},
but also in some other ``core-collapsed'' clusters, like NGC 6752  \citep{1997ApJ...474..701R},
the upper limit for the binary fraction can be as high as $\sim30\%$.

For the {\em initial\/} binary fraction
in globular clusters, there are of course no direct measurements.
However, there are no observational or theoretical arguments
suggesting that the formation of binaries and hierarchical multiples
in dense stellar systems should be significantly different from
other environments like open clusters, the Galactic field,
or star-forming regions.
Binary frequencies $\ga 50\%$ are found in the
solar neighborhood and in open clusters \citep{2003A&A...397..159H}.
T Tauri stars also have a very  high binary fraction \citep{2001AGM....18..P24K}.
For the range of separations between 120 and 1800 AU, their binary fraction is comparable to that of
main sequence  stars in the solar neighborhood \citep{1996A&A...307..121B},
while at shorter periods it is higher \citep{2003A&A...410..269M}.
Furthermore, many stars are formed in systems of multiplicity
3 or higher: in the field their abundance  is no less than 40\% 
for inner periods $\le 10$ days \citep{1997AstL...23..727T}.
All this suggests that, in dense stellar systems as well,
most stars could be formed in binary and multiple configurations.

Most dynamical interactions in dense cluster cores tend to {\em
destroy\/} binaries (the possible exception is tidal capture,
which may form binaries, but turns out to play a negligible role; see
\S~5.2).  Soft binaries (with orbital speeds lower than the cluster
velocity dispersion) can be disrupted easily by any strong encounter  with
another passing star or binary.  Even hard binaries can be destroyed
in resonant binary--binary encounters, which typically eject two
single stars and leave only one binary remaining
\citep{1983MNRAS.203.1107M}, or produce physical stellar collisions
and mergers \citep{1996MNRAS.281..830B,Fregeau_FB2_04}.

In addition, many binary stellar evolution processes lead to
disruptions (e.g., following a supernova explosion of one of the
stars) or mergers (e.g., following a common envelope phase).
These evolutionary destruction processes can also be enhanced by  dynamics.
For example, more common envelope systems form as a result of exchange
interactions \citep{2000ApJ...532L..47R}, and
the orbital shrinkage and the development of high eccentricities
through hardening encounters
may lead to the coalescence of binary components  \citep{1984AJ.....89.1811H, 2003ApJ...589..179H}.

It is therefore natural to ask whether the small binary fractions
measured in old globular clusters today result from these many
destruction processes, and what the {\em initial binary fraction\/}
must have been to explain the current numbers. We address these
questions in this paper by performing calculations that combine
binary star evolution with a treatment of dynamical interactions in
dense cluster cores.
In \S~3 we describe in detail the method we use, following a brief
overview of the theoretical background in \S~2.
We test our simplified dynamical model by comparing it against full  Monte Carlo
$N$-body simulations in \S~4.1. In \S~4.2 we use semi-analytical  estimates to predict
the upper limit for the final binary fraction in  dense clusters.
In \S~4.3 we estimate the lower limit for the final binary fraction
and analyse which mechanisms of binary destruction  are most efficient
as a function of cluster age. In \S~5
we present our numerical results for the evolution of the binary  fraction
in dense cluster cores, and we compare these results with observations.
In particular, using our theoretically predicted period distribution,
we re-examine observations of 47~Tuc and re-derive constraints
on the core binary fraction. In the final discussion (\S~6), we point  out
how our results may be helpful in interpreting observations of
core binary fractions in other clusters, and we discuss the required  initial
conditions
for simulations of clusters with binaries, as well as which methods are  best
suited for these simulations.

\section{Binary Population Synthesis with Dynamics}

There are several possible ways to approach the study of binary  evolution in
dense clusters.
The traditional approach is to start from full $N$-body simulations to  study
the dynamics of the stellar system and introduce on top of this various  simplified
treatments of single and binary star evolution.
This has been used for many years
\citep[for recent examples see][]
{2001MNRAS.321..199P,2002ApJ...571..830S,2003ApJ...589..179H}.
Unfortunately, even with the fastest special-purpose computers available
today, this direct $N$-body approach remains extremely expensive  computationally,
so that previous studies have been limited to
small systems like open clusters and with limited coverage of parameter  space.
In addition, because binaries are particularly expensive to handle  computationally
(as they increase enormously the dynamic range of direct $N$-body  simulations),
these previous studies have also been performed with unrealistically  small
numbers of binaries. For example, the
time required to perform just one direct $N$-body simulation of a  cluster
containing $2\times 10^5$\ stars 
with all stars formed initially in binaries would be at least a year
on the GRAPE-6, with some dependence on the initial binary parameters
\footnote{J.~Hurley, personal communication. The estimate is based on 5 days 
required to simulate an open cluster of 20000 stars
with 2000 binaries on the GRAPE-6 in \citet{2002ApJ...571..830S}}

Alternatively, a binary population synthesis code \citep[e.g.,][]{Hurley_Binary_02},
normally used to evolve large numbers of stars and binaries without
dynamical interactions, can be extended by
introducing a simple treatment of dynamics.
In this type of approach it is often assumed that all the relevant
parameters of the cluster
(e.g., central density and velocity dispersion)
remain constant throughout each dynamical simulation, i.e., the dynamics
is assumed to take place in a {\em fixed background\/} cluster.
Many previous studies of dense stellar
systems have been based on this type of approximation
\citep[see, e.g.,][]{hut_1992, 1994ApJ...437..733D, 1995ApJS...99..609S,
ecology_i, ecology_ii, 1995MNRAS.276..887D, 1995MNRAS.276..876D, 1997MNRAS.288..117D, 2000ApJ...532L..47R,  2001MNRAS.322L...1S}.
This approach, sometimes called ``encounter rate technique''  \citep{2002LRR.....5....2B},
is computationally much less expensive than direct $N$-body simulations
and hence allows the systematic exploration of the vast parameter space
of initial conditions for clusters and their primordial binary  populations.
In addition, the use of sufficiently large numbers of stars and binaries
makes the simulations more realistic.
Although obviously much less accurate in its description of the overall
cluster dynamics, this method
opens the possibility of studying ``star cluster ecology'' in  considerably
greater detail than has been possible with $N$-body simulations.
In particular, it makes it possible to study in detail the rare but  important
evolutionary channels that may play a crucial role in the formation of
some of the most interesting tracers of dynamical interactions in dense
clusters, such as ultracompact X-ray binaries, millisecond pulsars, and
cataclysmic variables \citep{Aspen_2004, 2004RMxAC..20...67I}.

Unfortunately, it is difficult to compare these two approaches, as each  is
based on a very different set of simplifying assumptions.
There are no comprehensive studies of dense stellar systems
including a self-consistent treatment of both dynamics and
binary star evolution.
In many recent $N$-body simulations for large clusters 
\citep[using either Aarseth-type codes or H{\'e}non's Monte Carlo method;][] 
{2001NewA....6..277A, 2003ApJ...593..772F},
binary stars are treated in the point mass limit and soft binaries are
eliminated from the start.
Binary destruction can then occur only through resonant 4-body  interactions.
However, $N$-body studies of open clusters that incorporate realistic  treatments
of binary stellar evolution have shown that
stellar evolution affects the binaries significantly, and that, even in
these low-density environments, the complex interplay between binary
evolution and dynamics, even for soft binaries, can play an important
role in the overall cluster evolution and in determining the properties
of surviving binaries \citep{2003ApJ...589..179H}.

The second approach, ``binary population synthesis with dynamics'',  which we
have adopted in this work, suffers
from the lack of self-consistent dynamical evolution of the cluster,
which is assumed to remain in a constant state of thermal equilibrium
for its entire evolution.
This state, where the energy production through ``binary burning'' in  the core
is balanced by the outer energy flux into the cluster halo, does
indeed provide  nearly constant conditions throughout a typical globular
cluster's lifetime. The exception might be ``core-collapsed'' clusters,  which may
have run out of binaries and evolved to a much more centrally  concentrated state.
Typically, the density and ``temperature'' profiles of a cluster do not  change much
as long as there are enough binaries remaining to provide support  against
gravothermal contraction.
Stellar interiors provide a useful analogy: as long as a star keeps  enough hydrogen
to burn in its core, it can remain in thermal equilibrium on the main  sequence
and avoid core contraction and envelope expansion.
Just like main-sequence stars, globular clusters can maintain
a nearly constant interior structure for many thermal time-scales
(i.e., many relaxation times) as long as they do not run out of ``fuel''
(binaries).
This behavior is expected qualitatively \citep[see,  e.g.,][]{GoodmanHut_89}, and has
now been demonstrated quantitatively in many different studies using  various
numerical techniques for cluster simulations.
For example, the recent study by \citet{2003ApJ...593..772F} considered
the evolution of idealized clusters of equal-mass stars without stellar evolution for
a range of initial binary fractions.
For the case of an isolated Plummer model with 10\% initial hard binaries they found 
(see their Fig.~4) that the core radius of this cluster can remain nearly constant 
(to within a factor $\sim 2$) for  many tens of
half-mass relaxation times (i.e., more than a Hubble time for most  Galactic
globular clusters, where the half-mass relaxation time is $ \sim 10^9$ yr).
For a more realistic cluster model with 20\% hard binaries initially and
tidal truncation, they found that, after about $40\,t_{\rm rh}$, when  the
cluster is about to disrupt in the
Galactic tidal field, the core radius still has not varied by more than  a
factor $\sim 2$ over the entire evolution (see their Fig.~11);
and over the first $10\,t_{\rm rh}$, the core radius
changed by less than $\sim20\%$.
The central velocity dispersion also does not vary
much with time \citep[see, e.g.,][Fig.~1]{2003MNRAS.343..781G}.
Similar results have been obtained from direct $N$-body simulations of  open
clusters, where the central density and velocity dispersion
also remain nearly constant in models with significant
numbers of binaries \citep{2003ApJ...589..179H}.

There are several binary population synthesis codes in use today.
Only a few of them include, in addition to detailed treatment of mass transfer phases, stellar evolution
along with tidal interaction of binary components.
The code of \citet{Hurley_Binary_02} was designed to study low- and intermediate-mass stars
leading to the formation of white dwarf systems.
The code of \citet[][in preparation]{Chris_02,Chris_04} was originally developed to
investigate more massive stars---progenitors of neutron star and
black hole systems (calibrated against full binary stellar evolution codes for 
mass-transfer phases)---and was expanded recently to treat carefully binaries with white dwarfs as well.
We use the latter code (called {\tt StarTrack}) to follow the evolution of single
and binary stars in our simulations.
As mentioned earlier, we also  treat all dynamical encounters explicitly by direct integration, using a 
recently developed numerical toolkit for small-$N$ gravitational
dynamics that is particularly well suited to performing 3-body and 4-body
integrations \citep{Fregeau_FB2_04}.

\section{Methods and Assumptions}

\subsection{Cluster Initial Conditions}

Our initial conditions are described by the following parameters:
total number of stars $N$ (single or in a binary),  initial mass function
(IMF), binary fraction $f_b$, distribution of binary parameters
(period, $P$, eccentricity, $e$, and mass ratio, $q=m_2/m_1 <1$). We typically
adopt standard choices used in previous population synthesis studies, which are
based on available observations for stars in the field and in young
star clusters (Sills et al.\ 2003).  For most of the calculations reported here, we
use the following ``standard'' initial conditions:

\begin{itemize}

\item  We adopt the IMF of \citet{Kroupa_IMF_02}, which can be written as a broken power law
$dN\propto m^{-\alpha}dm$, where $\alpha=0.3$ for $m/M_\odot<0.08$, $\alpha=1.3$ for $0.08\le m/M_\odot<0.5$,
$\alpha=2.3$ for $m/M_\odot \ge 0.5$. We assume that all stars are  formed
in a single burst of star formation at $t=0$ in our simulations.

\item We consider a wide range of stellar masses from 
$0.05\,M_\odot$  to $100\, M_\odot$\footnote{The lower limit is chosen in order to provide
a self-consistent mass-ratio distribution for binaries with primaries down to $0.15\ M_\odot$.}

\item The binary mass ratio, $q$, is assumed to be distributed uniformly
in the range $0 < q <1$. This is in agreement with observations for $q  \ga 0.2$
\citep{Woitas_MassRatio_01}.

\item The binary period, $P$, is taken from a
uniform distribution in $\log_{10} P$ over the range $P =
0.1$--$10^7\,$d.

\item The binary eccentricity, $e$, follows a thermal
distribution with probability density $p(e) = 2e$.

\item We reject systems where binary components would overflow their roche lobe at pericenter.

\end{itemize}

The initial average stellar mass is then $\langle m \rangle \simeq 0.5\,M_\odot$,
and, with the flat mass ratio distribution, the average binary mass is
$\langle m_{\rm b} \rangle \simeq 0.75\,M_\odot$.

\subsection{Stellar Evolution}

We evolve all stars (single and binary) using the population synthesis
code {\tt StarTrack\/} \citep{Chris_02}, which has
recently been updated significantly \citep[][in preparation]{Chris_04}.
This is the only current population synthesis code that
incorporates detailed treatments of all possible types
of mass transfer (MT) episodes: stable MT (conservative or  non-conservative),
unstable MT (thermally or dynamically), and thermal time-scale MT. Also  included
are the effects of Eddington-limited mass accretion
and transient behavior of accretion discs (based on the disc instability model).
{\tt StarTrack\/} allows us to follow the evolution of binaries with a  large range of
stellar masses, metallicities, and star formation histories
(constant rate, sudden or exponential bursts, etc.).
{\tt StarTrack} also models in detail the loss of mass and angular  momentum through
stellar winds (dependent on metallicity) and gravitational radiation,
asymmetric core collapse events with a realistic spectrum of compact  object masses,
and the effects of magnetic braking and tidal circularization on close  binaries.
In our simulations, we adopted
the new prescription for magnetic braking given by
\cite{2003ApJ...599..516I}. Compared to the older prescription \citep{1981A&A...100L...7V}
closer binaries lose their angular momentum at a slower rate and hence survive as  binaries longer.
The evolution of single stars in {\tt StarTrack\/} is based on the
analytic fits provided by \cite{Hurley_Single_00}, but
includes a more realistic determination of compact object masses
\citep{FryerKalogera_BH_01}.

We treat the evolution of stellar collision and binary merger products  following
the general ``rejuvenation'' prescription of \cite{Hurley_Binary_02}.
It ensures that the merger product has the same total amount of  hydrogen, helium, and
carbon as the two parent stars together.
For some stars, the assumptions made in the treatment of the merger
depend on the types of stars and the type of merger (collision vs binary
coalescence). For example,
we assume that there is no accretion on to a neutron star during a  physical collision,
and that the other star, if it is unevolved, is destroyed completely
(e.g., we do not consider the possible formation of a Thorne-\.Zhytkow object).
We treat as a ``dynamical common envelope'' (CE) event the outcome
of a physical collision between a compact object
and a red giant, applying a standard ``alpha prescription''
\citep[where we adopt $\alpha_{\rm CE}\lambda=1$; see][]{1993PASP..105.1373I},
but taking into account the initial positive energy.
In particular, if this compact object is a neutron star,
a compact binary containing a neutron star and a white dwarf is formed.
We assume that, during the CE phase, the neutron star will accrete a significant amount of
the envelope material and will become a millisecond pulsar  \citep{1998ApJ...506..780B}.
If the resulting mass of the  neutron star exceeds the limit for a  neutron star
(taken in our simulations to be $2 M_\odot$), we assume that a  black hole is formed.

To evolve the cluster population of single stars and binaries in our  code,
we consider two basic time-scales.
One is associated with the evolutionary changes in the stellar  population,
$\Delta t_{\rm ev}$, and the other with the rate of encounters,
$\Delta t_{\rm coll}$ (see \S~3.4).
The evolution timestep $\Delta t_{\rm ev}$ is computed so that no more
than 2\% of all
stars change their properties (mass and radius) by more than 5\%.
The global timestep for the cluster evolution is taken to be $\Delta t=
\min [t_{\rm ev}, t_{\rm dyn}]$

\subsection{Dynamical Cluster Model}

As we described in \S~2, our model for the cluster dynamics is highly  simplified.
We adopt a simple two-zone, core-halo model for the cluster.
We assume that the core number density, $n_{\rm c}$, and  one-dimensional
velocity dispersion, $\sigma_{\rm 1D}$, as well as the
half-mass relaxation time, $t_{\rm rh}$,
remain strictly constant throughout the evolution.
While dense globular clusters of interest have $\sigma_{\rm 1D} \sim 10\,{\rm  km}\,{\rm s}^{-1}$, the
core density can vary by several orders of magnitude. Here we set
$n_{\rm c}=10^5\,{\rm pc}^{-3}$ for most calculations, representative
of a fairly dense cluster like 47~Tuc. In general, $n_{\rm c}$ is the
main ``knob'' that we can turn to increase or decrease the importance
of dynamics.  Setting $n_{\rm c}=0$ corresponds to a traditional
population synthesis simulation, where all binaries and single stars
evolve in isolation after a single initial burst of star formation.
To model a specific cluster, we match its core mass
density today, $\rho^{\rm ob}_M$,
central velocity dispersion, and half-mass relaxation time.

The escape speed from the cluster core can be estimated from
observations as $v_{\rm e} \simeq 2.5\,\sigma_{\rm  3D}$ \citep{Webbink_GC_90}, where
$\sigma_{\rm 3D}$ is the three-dimensional central velocity dispersion.
Following an interaction or a supernova explosion, any object that has
acquired a recoil speed exceeding $v_{\rm e}$ is removed from the
simulation. Acquiring a large recoil velocity in a dynamical encounter
is a very efficient mechanism for ejecting low-mass objects from the  cluster.
We find generally that recoil to the halo does not play a significant role: the recoil velocity 
into the halo differs by $\sim 10\%$ from the escape velocity from the cluster, and affects only a small number of objects.

For computing interactions in the core, the velocities of
all objects are assumed to be distributed according to a lowered
Maxwellian \citep{King_65}, with
\begin{equation}
f(v) = v^2/\sigma^2(m)
  \left[ \exp(-1.5 v^2/\sigma^2(m))  - \exp(-1.5  v_e^2/\sigma^2(m))\right],
\end{equation}
where $\sigma(m)= (\langle m \rangle_{\rm c}/m)^{1/2}\sigma_{\rm 3D}$  (assuming energy
equipartition in the core) and $v_{\rm e}$ is the escape speed.
Here $\langle m \rangle_{\rm c}$ is the average mass of an object in  the core.
In addition, we use $\sigma$ to impose a cut-off for soft binaries  entering the core:
Any binary with maximum orbital speed $< 0.1 \sigma_{\rm 3D}$ is immediately
broken into two single stars \citep{Hills_90}.

In the presence of a broad mass spectrum, the cluster core is always
dominated by the most massive objects in the cluster, which tend to
concentrate there via mass segregation.  As stars evolve, the {\em
composition\/} of the core will therefore change significantly over
time. Mass  segregation in globular clusters
was investigated recently in \citet{Fregeau_MS_02} by considering
both light and heavy tracers in two-component models.
It was found that the characteristic mass-segregation time-scale is  given by
\begin{equation}
\label{t_sc}
t_{sc}\simeq 10\ C \left(\langle m \rangle_{\rm h}/m\right) t_{\rm rh}  \ .
\end{equation}
Here $C$ is a constant of order unity and $\langle m \rangle_{\rm h}$
is the average mass of an object in the halo, and $m$ is the current mass of the object.
This equation represents a diffusion process and can be applied to all  stars,
not just to those more massive than average. Indeed, even low-mass  objects
may (rarely) diffuse into the cluster core on a long time-scale,  although on
average they will tend to drift outwards.
However, for very light objects with masses $\la 0.4 \langle m  \rangle_{\rm h}$
 (which is typically $\sim0.3\,M_\odot$ at the beginning and 
$\sim0.15\,M_\odot$  after $\sim 10\,$Gyr),
this expression becomes less accurate, although it remains valid  qualitatively.
For a more recent discussion of mass segregation in the presence of a  broad
mass spectrum, see \cite{2004ApJ...604..632G}.

To model mass segregation in our simulations, we adopted the time-scale
given by equation~(\ref{t_sc}), but treated the process as stochastic:
the probability for an object of mass $m$ to enter the core after a time
$t_s$ is sampled from a Poisson distribution,
\begin{equation}
\label{pts}
p(t_s)=(1/t_{sc})\exp(-t_s/t_{sc}).
\end{equation}
This treatment ensures that all stars heavier than
$\sim 0.4 \langle m \rangle_{\rm h}$
diffusing into the core will have the appropriate mass spectrum
and that interactions will occur between objects drawn from the
correct distribution.

Eq.~(\ref{t_sc}) was derived for the restricted case of a two-component 
cluster -- without a realistic IMF -- therefore $C$ is unknown by a factor of a few.
We find in simulations that the final core mass is nearly proportional to $1/C\langle m \rangle_{\rm  h}$.
In order to obtain a better fit to observations for the core mass
versus total cluster mass relation, we have fine-tuned eq.~(\ref{t_sc})
using data for 47~Tuc, in particular the ratio of the core mass to the total mass
of the cluster.
For this cluster we adopted a core mass of $10^5  M_\odot$
and a total cluster mass of $10^6 M_\odot$ at present  \citep{2001MNRAS.326..901F};
we also take $t_{\rm rh}=10^9$ yr \citep{Harris_GCcatalog_96} and
an age of $11\,$Gyr \citep{2003A&A...408..529G}.
While the core mass can be found directly from our simulations, the total 
cluster mass has an uncertainty due to the IMF cut-off at the low mass end
in our standard cluster model.
First, we found the total mass of a cluster model evolved to $11\,$Gyr
and the initial number of very massive stars (defined as those producing
a black hole at the end of their stellar evolution).
We find that at 11 Gyr, the cluster has $145 M_\odot$ per black hole
(or per heavy primordial star), when the IMF extends down to 0.01 $M_\odot$.
This allows us to normalize our model to the real cluster mass:
with this ratio we have an estimate for the total cluster mass when  we use
a higher cut-off for the IMF (0.05 $M_\odot$).
This now gives us the ratio of the core mass to the total cluster mass corresponding to our simulations.
We find that the best fit for 47 Tuc gives $C\langle m \rangle_{\rm  h}=3 M_\odot$.

\subsection{Treatment of Dynamical Interactions}

All objects in our simulations are allowed to have dynamical  interactions
only after they have entered the cluster core. We use a simple Monte Carlo prescription
to decide which pair of objects actually have an interaction
during each timestep.

The cross section for an encounter between two objects, of masses
$m_i$ and $m_J$, with relative velocity at infinity $v_{iJ}$, is
computed as
\begin{equation}
S_{iJ} = \pi d_{max}^2 (1+v_{p}^2 / v_{iJ}^2)\ ,
\end{equation}
where $d_{max}$ is the maximum distance of closest approach
that defines a significant encounter
and $v_{p}^2 = 2G (m_i + m_j)/d_{max}$ is the velocity at
pericenter. Here the index $i$ (lowercase) reflects an individual
object in the core, while $J$ (uppercase) denotes a random
representative object from the {\em subclass\/} of objects $J$. 
In order to more accurately determine encounter rates, 
at each timestep binaries and single stars in the core are divided into 100 subclasses: 
10 by size (radius for single stars or semimajor axis for binaries) and 10 by mass. 
Boundaries between mass-subclasses are fixed approximately as $0.1 \times 2^n$.
Subclasses by size depend on the current sizes of single and binary populations (separately), with
the step between subclasses taken as
$\delta\log_{10} r_{\rm bin} = 0.1 (\log_{10} R_{\rm max} -\log_{10} R_{\rm min})$. 
The encounter rate for a given object $i$
and an object from subclass $J$ is $\Gamma_{i,J} = n_J S_{iJ} v_{iJ}$,
where $n_J$ is the number density of objects in subclass $J$, and the
cross section and relative velocity are defined for an average object
in subclass $J$.

The total interaction rate for a given object $i$ is the sum of the
interaction rates with all relevant subclasses, $\Gamma_{i}=\sum_J n_J
S_{iJ} v_{iJ}$. The corresponding interaction time is $\tau_{i} =
1/\Gamma_i$. The actual time for an encounter $t_i$ follows a Poisson
distribution with mean $\tau_{i}$. In practice, we generate a random
number $0<X<1$, and assume that the encounter happened if
$t_i=X\tau_{i}\le \Delta t$. The timestep is limited so that $\Delta
t_{\rm dyn } \le 0.25 \min_i \tau_i$. We keep track separately of the
time-scales $\tau_{i,J}$ for interactions with each subclass $J$, and
the corresponding $t_{i,J}=X_J\tau_{i,J}$ is generated from an independent random
number. If an encounter happened, it is assumed to be with an object
from the subclass with the smallest $t_{i,J}$. The actual interacting
object $\hat J$ from that subclass $J$ is randomly selected from the
list of non-interacted objects in this subclass.

In this paper we consider separately binary--binary, binary--single, and
single--single interactions. The cross sections and rates are
calculated using $d_{max} = 5 (R_i + \langle R\rangle_J)$ for
single--single, $d_{max} = 3 (b_i+\langle R \rangle_J)$ for
binary--single and $d_{max} = 3 (b_i+\langle b \rangle_J)$ for
binary--binary.  Here $b_i=a_i(1+e_i)$ is the apocenter separation of
the binary ($a_i$ is the semi-major axis and $e_i$ is the
eccentricity), $\langle R \rangle_J$ is the average stellar radius in
the subclass $J$, and $\langle b \rangle_J$ is the average apocenter
separation in the subclass $J$.  A single--single interaction with
pericenter distance $d_i \le 2(R_i +R_J)$ is treated as a physical
collision and assumed to lead to a merger or a dynamical CE phase.
If $2 < d_i / (R_i + R_J) < 5$, we check whether a binary could be  produced via tidal
capture using the approach described in \cite{Zwart_TC_93}.  If a
tidal-capture binary is formed, its eccentricity is set to zero and
semi-major axis set to $2d_i$, assuming rapid circularization ($\sim 10$ yr) as
predicted by the standard model described in \citet{1987ApJ...318..261M}.

Each dynamical interaction involving a binary is calculated using \Fewbody,
a new numerical toolkit for simulating small-$N$ gravitational
dynamics that provides automatic calculation termination and
classification of outcomes \citep[for a detailed description see][]{Fregeau_FB2_04}.
\Fewbody\ numerically integrates the orbits of small-$N$ systems, and performs
collisions in the sticky-star approximation.  \Fewbody's
ability to automatically classify and terminate calculations as soon
as the outcome is unambiguous makes it well-suited for carrying out 
large sets of binary interactions, for which calculations must be as 
computationally efficient as possible.

\section{Test Calculations and Simple Estimates}

\subsection{Comparison with $N$-body Simulations}

We have compared our simple dynamical model to recent results from
fully self-consistent Monte Carlo simulations based on
H\'enon's algorithm for solving the Fokker-Planck equation  \citep{2000ApJ...540..969J,2001ApJ...550..691J}.
For our test we used the results obtained for
an idealized model cluster containing 20\% primordial hard binaries
(binding energies in the range $1 - 133$~$kT$, where $kT$ is the average  kinetic
energy of an object in the cluster)
for a King model with dimensionless central potential $W_0=7$
\citep[][hereafter F03]{2003ApJ...593..772F}.
In this simulation all stars had equal masses, were treated as point  masses
(no physical collisions), no stellar evolution was taken into  account,
and all binary interactions were treated using simple recipes.

We used our code to perform a similar simulation: we considered a  cluster
consisting of equal-mass stars, and we turned off all stellar
evolution and physical collisions.
To fit the F03 model, we took the core mass as 8.3\% of the total  cluster mass
(corresponding to a King model with $W_0=7$), and
we took an average star mass of $1\,M_\odot$ and  $\sigma_{\rm  r}=10\,$km/s.
For a total cluster mass of $3\times 10^5\,M_\odot$ these conditions  imply
$t_{\rm rh}=8\times10^8\,$yr, and $n_{\rm c}=2000\,{\rm pc}^{-3}$.
We evolved the cluster for $20\,t_{\rm rh}$.

\begin{figure}
\includegraphics[width=84mm]{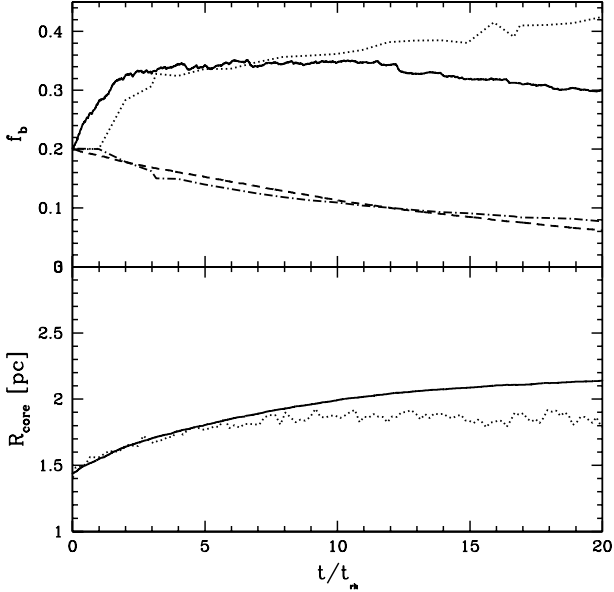}
\caption{Evolution of the core and halo binary fractions (top) and the core radius (bottom) in our test model,
compared with the F03 model (see text).
In the top panel, the solid line shows the binary fraction in the core and the dashed  line shows
the binary fraction in the halo, both for the test model. 
The dotted line shows the binary fraction in the core in the F03 model
and the dash-dotted line shows the binary fraction in the halo in F03 model.
In the bottom panel the solid line shows the core radius in the test model and the dotted line
shows the core radius in the F03 model.}
\label{test_sf}
\end{figure}

In Fig.~\ref{test_sf} we show our results for the core and halo binary fractions
as a function of time, compared with the model of F03.  The agreement with F03  is  excellent:
the core binary fraction rises very quickly to $\sim 35\%$  and then  remains close
to this value for $\sim 10 t_{\rm rh}$.
In contrast, the halo binary fraction decreases more gradually
from 20\% to 10\%.
Considering the differences between the two treatments,
especially for binary interactions,
this agreement is quite remarkable.

In the bottom panel of Fig.~\ref{test_sf} we show the evolution of the core  radius $r_{\rm c}$.
The core radius evolves in the same way as in the
dynamical simulation of F03.
Overall, the core radius does not change much during the entire  evolution
and its value is consistent with the measured values for observed  globular clusters
with similar parameters (e.g., NGC 3201 or NGC 6254, in which
the total cluster mass, the central number density, and the central
velocity dispersion are similar to those in our model).

\subsection{Semi-analytic Estimates}

\label{sec:semianalytic}

\begin{figure}
\includegraphics[width=84mm]{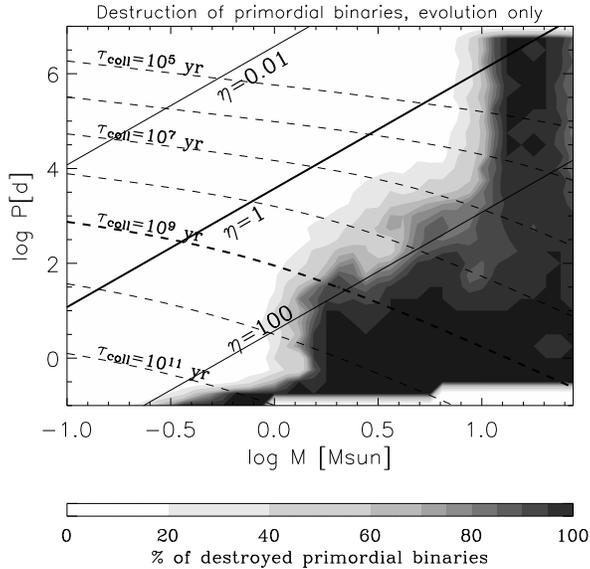}
\caption{Destruction of primordial binaries by stellar evolution, shown
in the parameter space of total initial binary mass and initial binary period.
Solid lines are lines of constant binary hardness and dashed lines are lines of
constant collision time.}
\label{fate_2d}
\end{figure}

One can estimate the final binary fraction $f_{\rm b,c}$ in a dense  environment
by considering several mechanisms of binary destruction:
\begin{itemize}
\item all soft binaries are usually destroyed during a strong encounter.
\item some fraction of hard binaries is destroyed through stellar  evolution
(mergers or disruptions after supernova explosions)
\item when a hard binary has a strong encounter with another hard binary or a single star,
it can exchange its less massive component for a more massive star,
shrink its orbit, or be destroyed in a collisional merger.
\end{itemize}

\noindent It should be noted that in our simplified semi-analytical treatment we neglect the
effect of mass segregation, which tends to increase the core binary fraction.  (This issue
is discussed in more detail in Section~\ref{sec:mainrefmodel}.)

Let us consider a dense environment with number density $10^5\,{\rm  pc}^{-3}$,
$\sigma_{\rm 1D} = 10\,$km/s and with an average mass of $0.5\,M_\odot$.
With our choices of initial parameters, and an
average mass of $0.5 M_\odot$,  40\% of all primordial binaries  initially are soft
(this fraction would be 50\% with respect to the average mass of $1 M_\odot$).
We introduce $\eta$ -- the hardness of a binary system--- as
\begin{equation}
\label{eta_def}
\eta = {\frac {G m_1 m_2} {a \sigma^2 \langle m\rangle }}\ ,
\end{equation}
\noindent where $a$ is the binary separation, $m_1$ and $m_2$ are the masses of the binary components,
and $\langle  m\rangle$ is the average mass of a single star.
Binaries that have $\eta < 1$ are termed soft, and those with $\eta > 1$ are termed hard.

To find how many hard binaries will be destroyed by stellar evolution  alone,
we calculated
the probability of binary destruction as a function of its initial  total mass
and orbital period, using the binary population synthesis code (see  Fig~\ref{fate_2d}).
This simulation was done with $1.25\times 10^5$ binaries
distributed initially flat in $\log P_{\rm d}$ and $\log M_{\rm tot}$ (in order to have
better resolution for destruction rates for high mass binaries), where
$P_{\rm d}$ is the binary period in days and $ M_{\rm tot}=M_1+M_2$ is  the total binary mass in $M_\odot$,
and was evolved for 14 Gyr.

The result is striking: most of the very hard binaries, with hardness
ratios $\eta \ga 100$, are destroyed by stellar evolution.
The empty space near the bottom right corner of Fig~\ref{fate_2d}  reflects the
absence of systems below the minimum period for binaries
on the main sequence (the period at which stars come into contact).
For binaries with total mass $\ge 10\,M_\odot$ destructions mainly occur
during the first $10^8$ years of cluster evolution.  Binaries with period  $\ge 10^4$ days
are mainly destroyed through SN explosions.  Binaries with period $\le 10$ days
are destroyed mainly via mergers at the MS stage.
For periods $10$--$10^4$ days the destructions
are associated with common envelope evolution and occur at later times.
Destructions in binaries of smaller masses are not much different from more massive binaries 
at small periods, but are not destroyed through SN explosions at large periods.
Also, the CE event in less massive binaries occurs when the donor is a less evolved giant
(at the first red giant branch).

One may expect that destruction rates should vary smoothly; however,
binary evolution involves many qualitatively different events. 
In particular, an interesting fluctuation in destruction rates can be seen at $\log P\sim 1.75$ and $\log M_{\rm tot} \sim 0.95$,
where local destruction rates are lower than for nearby binaries.
For these and nearby binaries the destruction rates are about the same for mass ratio $q\la 0.5$
but different for larger $q$.
For binaries of masses close to and smaller than these, most destructions for $q\ga 0.5$ occur during the
CE event between a WD and a giant. This CE phase is the second interaction in the binary and 
follows the stable MT event with the other donor.
When the total binary mass is smaller, the WD mass is $\la 0.9\, M_\odot$, and CE event leads to a merger.
For binaries of higher total mass, the second interaction occurs between a He star or a WD 
and a star at the Hertzsprung gap. This MT is unstable and leads to a merger. 
This, therefore, provides for a small local decrease in destruction rates.

With our IMF and considering binaries of all masses,
the fraction of hard binaries  destroyed during their evolution is 20\%.
Among  binaries where at least one star is more massive than  $0.5\,M_\odot$
the destruction fraction is 50\%, and for those with total initial
mass above $1\,M_\odot$, this fraction is closer to 60\%.
 The fraction of hard binaries that is not destroyed but is instead 
softened by evolution is very small, $\le 1\%$.
In \S4.3 we will discuss in more detail how binary destruction rates
change with time.

Therefore, even before any dynamical processes are considered for the
hard binary population, we estimate that the final binary fraction
cannot be higher than about 50\%, and for binaries with at least one star
more massive than $0.5\,M_\odot$ this upper limit becomes 30\%. For relatively
massive binaries, with total initial mass above $1\,M_\odot$,
the upper limit for $f_{\rm b,c}$ is only 24\%.
Overall, this estimate already shows that (i) the expected final binary  fraction
in a dense star cluster will be low and
(ii) stellar evolution cannot be neglected when estimating binary  fractions
from dynamical models of dense star clusters.

Let us now consider the effects of dynamical interactions.
The time-scale for a binary to undergo a strong encounter with another
single star, the collision time, can be estimated as  $\tau_{\rm  coll}=1/n\sigma v_\infty$.
Assuming that the strong encounter occurs when the distance of closest  approach
$d_{max}\le k a$ with $k\simeq 2$, we obtain

\begin{eqnarray}
\label{tcoll_pd}
\tau_{\rm coll} = 3.4 \times 10^{13} \ {\rm yr} \ \  k^{-2} P_{\rm  d}^{-4/3} M_{\rm tot}^{-2/3} n_5^{-1}
v_{10}^{-1} \times \\ \nonumber
\left( 1+913 {\frac { (M_{\rm tot} + \langle M\rangle )} {k P_{\rm d}^{2/3} M_{\rm  tot}^{1/3} v_{10}^2}}\right) ^{-1} 
\end{eqnarray}

Here $\langle M\rangle$ is the mass of an average single star in  $M_\odot$,
$v_{10}=v_{\infty}/(10\,{\rm km/s})$, where $v_\infty$ is the relative  velocity
at infinity and $n_5=n/(10^5\,{\rm pc}^{-3})$, where $n$ is the number  density.
Using eq.~(\ref{eta_def}), 
eq.~(\ref{tcoll_pd}) can be rewritten in a more convenient form:
\begin{eqnarray}
\label{coll}
\tau_{\rm coll}= 1.7 \times 10^{8} \ {\rm yr} \ \ \eta^2  k^{-2}  n_5^{-1}
{\frac {\langle M\rangle^2} {M_1^2M_2^2}} \\ \nonumber
\left(1+\eta {\frac {2} {k}} {\frac {M_{\rm tot}+\langle M\rangle} {M_1  M_2}} \langle M\rangle \right) ^{-1}
\end{eqnarray}

It can be seen from eq.~(\ref{coll}) that the collision time for
a binary with $\eta = 1$ and $M_{\rm tot}= 1\,M_\odot$  is only $\sim  10^8\,$yr.
Overall, with our IMF, $\sim 50\%$ of all hard binaries
have collision times shorter than $\sim10\,$Gyr (see also  Fig.~\ref{fate_2d}),
and 15\% have collision times as short  as $\sim1\,$Gyr.
In addition, most of the hard binaries that could have experienced an  encounter
are binaries with $\eta=1-100$, and therefore they are from the population  that is
destroyed by stellar evolution to a lesser degree than binaries harder than $\eta=100$.
While this estimate considered only binary--single encounters,
binary--binary encounters will further enhance the destruction rate.
Moreover, when the binary fraction is higher than $\sim 25\%$,  binary--binary encounters
dominate over binary--single encounters.

Each binary--single encounter with a hard binary can result in a hardening of this binary,
an exchange of a companion with a more massive single star, or
binary destruction in a physical collision.
The probability of a physical collision during an encounter increases  strongly as
the binary becomes harder \citep{Fregeau_FB2_04}.
In addition, a very hard binary can be ejected from the core if its
recoil velocity
exceeds the escape speed from the cluster.
Each of these three processes, directly or indirectly, leads to the
depletion of binaries (immediate or delayed):
acquiring a more massive companion, as well as orbital shrinkage
or the eccentricity increase,
makes systems more likely be destroyed through stellar evolution.
As a conservative lower limit, we assume
that half of the interacting hard binaries will be destroyed as a  result of an encounter (immediately or later).
This is clearly a lower limit, as scattering experiments show that,
for hard binaries containing main-sequence stars
most encounters will lead to a physical collision  \citep{Fregeau_FB2_04}.

Taking everything into account, we conclude that 64\% of all binaries  will
be destroyed --
only $k_{\rm s}=36\%$ of primordial binaries can survive to the present.
The expected final binary fraction is therefore
$f_{\rm b}=N_{\rm b}/(N_{\rm s}+N_{\rm b}) = k_{\rm s}/(2-k_{\rm  s})=22\%$
(for stars in the entire mass range), and it is 13\% ($k_{\rm s}=23\%$)
for binaries with at least one star more massive than $0.5\,M_\odot$,
while for $M_{\rm tot} \ge 1 M\odot$ it is only 10\% ($k_{\rm s}=18\%$).
We note again that this upper limit for the final binary fraction
does not take into account
several other  mechanisms leading to binary destruction, such as  binary--binary
encounters, which are more likely to cause binary destruction than
binary--single encounters.

\subsection{Encounters in Dense Environments}

\begin{figure}
\includegraphics[width=84mm]{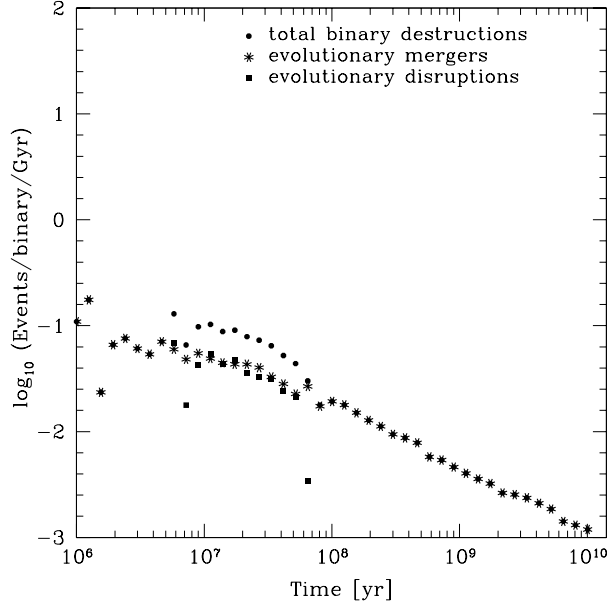}
\caption{Binary destruction rates as a function of time for a field  population
(i.e., without dynamical interactions).
Rates are given as numbers of events per binary per Gyr for mergers and
disruptions (following a supernova explosion).  Note the peak in evolutionary disruptions at $t\sim10^7$--$10^8$ yr,
which is due to supernovae.}
\label{dr_field}
\end{figure}

\begin{figure}
\includegraphics[width=84mm]{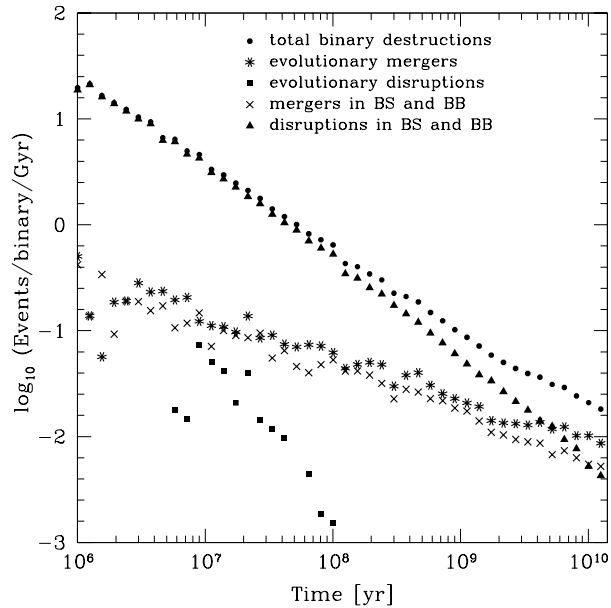}
\caption{Same as Fig.~\ref{dr_field}, but now for the same binary  population
evolved in a high-density environment with $10^5$ binaries per $\mbox{pc}^3$.  Here
disruptions through binary--binary (BB) and binary--single (BS)  interactions
dominate for the first $\sim 5$ Gyr.}
\label{dr_core}
\end{figure}

In order to verify our encounter rates,
we considered the evolution of a binary population that is completely
immersed into a dense environment with $n_c=10^5\ pc^{-3}$ for its
entire evolution. We adopted a velocity dispersion $\sigma_{\rm 1D} = 10$ km/s and evolved
the system for 14 Gyr.
This is an even simpler model of a star cluster,
where all objects are effectively inside the ``core'' at all times
and the effects of mass segregation and diffusion between a core and
a halo are completely ignored. It allows us to study more clearly the  various
types of dynamical interactions as separate processes.
We ran a sequence of simulations, with (i) only binary--single
encounters, with no physical collisions allowed during an encounter;
(ii) binary--single encounters with physical collisions;
(iii) both binary--single and binary--binary encounters but still  without
physical collisions; (iv)  binary--single and binary--binary encounters
with physical collisions;  and (v) all types of encounters allowed,
including single--single collisions. In this last case
we also eliminated from the system any object that acquired
a recoil speed exceeding the escape velocity $v_e = 2.5\, \sigma_{\rm 3D}$.

In \S~\ref{sec:semianalytic} we estimated that, with only  binary--single encounters
taken into account, the final binary fraction should not exceed $22\%$.
This assumed that all objects in the core had a  mass equal to the average mass $0.5\,M_\odot$.
With the complete IMF, a single star participating in the encounter can  have mass higher
than $0.5\,M_\odot$.  Also, a fraction of initially hard binaries can  become soft
during the evolution, e.g., because of mass loss. As a result, we  obtain even
lower remaining binary fractions: $f_{\rm b,c}=16.5\%$ for case~(i) and $f_{\rm b,c}=16.4\%$ for case~(ii). These
numbers are in agreement with the upper limits we estimated
in \S~\ref{sec:semianalytic}.

When binary--binary encounters are taken into account, the result is
even more dramatic: we obtain
$f_{\rm b,c}=8.0\%$ and $f_{\rm b,c}=7.9\%$
for cases~(iii) and~(iv), respectively.
As expected, we see that binary--binary encounters further enhance the
binary destruction rate.

When all dynamical processes are taken into account, in case~(v), we  obtain
a final binary fraction $f_{\rm b,c}=6.9\%$.
While our semi-analytical estimate of the binary fraction provided an
upper limit, the result obtained in this section
is clearly a lower limit: in a real cluster, not all binaries will be
exposed to the high interaction rates of the core at all times.
However, since more massive objects are  more likely to diffuse into  the core,
and since their destruction rate can be higher than for objects of  average mass,
this lower limit is only approximate, but we may expect it to be much  closer
to the real value than our previous, conservative upper limit.

Let us now consider in detail which mechanisms of binary destruction are
most important.
In Fig.~\ref{dr_field} we show the rates of binary destruction (events
per binary per Gyr) for a field population (with no interactions).
Except for the interval between $\sim 10^7$ and $\sim 10^8$ years, when  black holes
and neutron stars are formed, the binary destructions are mainly
coming from mergers. 
The power-law behavior observed for the overall  destruction rate at late times
is on the one hand imposed by the rate of orbital decay driven by magnetic braking,
and on the other hand depends on the evolutionary time-scale for the companion to become a red giant.

In Fig.~\ref{dr_core} we show the binary destruction rates again, but  for the
dense environment.  During the first few Gyr the
main destruction mechanism is the dynamical disruption of soft binaries.
The rate of disruption through stellar evolution is smaller than in the  field
population: some massive binaries are destroyed by  dynamical encounters
before they evolve to a supernova explosion.  However, the rate of
evolutionary mergers is about the same as for the field for the first $\sim10^8$ years.
This is consistent with our estimate for the collision
time of hard binaries: at this time, hard binaries
have not yet been hardened significantly by dynamical encounters.
After  $\sim 10^8\,$yr
the rate of binary evolutionary mergers is increased compared to
the field population, as the dynamical hardening of hard binaries
has now started.  The rates
for binary destruction through mergers and physical collisions are very  similar:
in most cases, if a binary is tight enough for an evolutionary merger,
the most likely outcome of a dynamical interaction is a physical  collision.

At about 5 Gyr, the rate of binary destruction through physical  collisions
starts dominating over dynamical disruptions: at this stage, almost no
soft binaries are left, and the hard binaries are more likely to
lead to physical collisions during an encounter. Consider a binary
with $\eta \sim 1$ at this stage. From equation~(\ref{coll}) we
can find that a collision time of $5\,$Gyr
corresponds typically to
binary components of $0.1\,M_\odot$ and $0.05\,M_\odot$
(assuming an average mass ratio of 0.5 and average field mass as  $0.5\,M_\odot$).
This binary is clearly at the lowest-mass end of our IMF, and therefore  by this
time almost all soft binaries in the simulation have been destroyed.
On the other hand, consider some more massive binary, which
has evolved through about half of its main-sequence lifetime,
and has component masses of $0.8\,M_\odot$ and $0.4\,M_\odot$ (and
still the same collision time of 5 Gyr).
The hardness of such a binary is $\eta\simeq 100$, corresponding to
a binary separation $\simeq 10\,R_\odot$.
For such a tight binary, the probability of a physical collision in an  encounter
is almost 100\%. The total rate of binary destructions through physical collisions
become comparable to that of dynamical disruptions, as can be seen from Fig.~\ref{dr_core}.

\section{Results}

\subsection{Overview of Cluster Models}

\begin{table}
\caption{Initial Conditions for All Models}
\label{table1}
\begin{tabular}{@{}l l l l l l l}
\hline
  Model                       & $\log n_{\rm c}$ 
& $\log \rho^{\rm ob}_{ M}$   & $\log t_{\rm rh}$
& $f_{\rm b,0}$               & $\sigma_{\rm 1D}$
& $v_{\rm e}$                   \\
\hline
 & & & & & &  \\
 1    & 5.0 &  & 9.0  & 1.0 & 10.& 43.\\
 D3   & 3.0 &  & 9.0  & 1.0 & 10.& 43.\\
 D4   & 4.0 &  & 9.0  & 1.0 & 10.& 43.\\
 D6   & 6.0 &  & 9.0  & 1.0 & 10.& 43.\\
 T8   & 5.0 &  & 8.0  & 1.0 & 10.& 43.\\
 T10  & 5.0 &  & 10.  & 1.0 & 10.& 43.\\
 B05  & 5.0 &  & 9.0  & 0.5 & 10.& 43.\\
& & & & & \\
M12     & 3.8 & 3.5$^{\rm a}$
                  & 9.02  & 1.0 & 4.5  & 19.6 \\
M4       & 4.4 & 4.1 &8.82  & 1.0 & 4.2 &20.3  \\
47~Tuc    & 5.3 & 5.1 & 9.48  & 1.0 & 11.5  &56.8\\
NGC~6388  & 5.9 & 5.7 & 9.08  & 1.0 & 18.9   & 78.2 \\
\hline
\end{tabular}

\medskip
Notations: $n_{\rm c}$ is the core number density in  ${\rm pc}^{-3}$ (assumed fixed),
$\rho^{\rm ob}_M$ is the observed core mass density in $M_\odot\,{\rm  pc}^{-3}$,
$t_{\rm rh}$ is the half-mass relaxation time in yr,
$f_{\rm b,0}$ is the initial binary fraction,
$\sigma_{\rm 1D}$ is the 1-D velocity dispersion in ${\rm km}\,{\rm  s}^{-1}$, and
$v_{\rm e}$ is the escape speed in ${\rm km}\,{\rm s}^{-1}$ \\

\medskip
$^{\rm a}$ $t_{\rm rh}$
    for specific clusters are taken from \cite{Harris_GCcatalog_96},
    $\rho^{\rm ob}_M$ and $\sigma_{\rm 1D}$  from \cite{Pryor_GCdisp_93},  and $v_{\rm e}$ from \cite{Webbink_GC_90}.
\end{table}

Initial conditions for all our models are given in Table~1.
The first group of models is used to cover the parameter space of
initial conditions over fairly wide ranges.
Our main reference model, Model~1, has
core density $n_{\rm c}=10^5\,{\rm pc}^{-3}$,
half-mass relaxation time $t_{\rm rh}=10^9\,$yr,
initial binary fraction $f_{b,0}=1$,
central velocity dispersion $\sigma_{\rm 1D}=10\,{\rm km}\,{\rm s}^{-1}$,
and central escape speed $v_{\rm e} = 2.5 \sigma_{\rm 3D} =
43\,{\rm km}\,{\rm s}^{-1}$.
We then consider three models with different central densities (D3, D4,  and D6),
two with different half-mass relaxation time (T8 and T10)
and one with an initial binary fraction decreased to 50\% (B05).
The initial total number of stars is $N = 2.5\times 10^5$ for all  models,
except for the ``47~Tuc'' model (see below), where
we used  $N = 5\times 10^5$). In all these models
the cluster core was assumed to contain about 1\% of the stars  initially,
and the metallicity was fixed at $Z=0.001$ (these two parameters have
very little influence on our results).

We performed several simulations with parameters that attempt
to match those of specific globular clusters in the Galaxy (the bottom  part of Table~1).
All these clusters are classified
observationally as "non-core-collapsed", meaning that they are well  fitted
by standard King models. These are precisely the kinds of clusters that,
theoretically, we expect to be in the ``binary burning'' thermal  equilibrium
state. For this set, we tried to consider the maximum range of
dynamical parameters, while concentrating on clusters at relatively
small distances and/or very well studied observationally, so that  current
or future observations could test our predictions.
For our most important model, 47~Tuc,
we also considered additional models in which we varied the initial  binary fraction
or introduced a time-dependent core density (See \S~5.5).
For all our models of specific observed globular clusters, the central
number density was chosen in order to provide (at the actual age of
the cluster, $\sim 11 - 13$\,Gyr) the best fit to the
observed mass density $\rho^{\rm ob}_M$ \citep{Pryor_GCdisp_93}.
Metallicities for these models are taken from  \cite{Harris_GCcatalog_96}.

\begin{table}
\caption{Results for Reference Models}
\label{table2}
\begin{tabular}{@{}l l l l l l}
 \hline
 Model	 
	& $\log \rho_{M}$   & $f_{\rm b,c}$    
	& $f_{0.5}$         & $f_{\rm wd}$ \\
\hline
 1    & 4.7    & 0.095   &  0.15   & 0.080 \\
 D3   & 2.7    & 0.265   &  0.37   & 0.165 \\
 D4   & 3.7    & 0.170   &  0.25   & 0.115 \\
 D6   & 5.7    & 0.035   &  0.055  & 0.055 \\
 T8   & 4.5    & 0.11    &  0.13   & 0.085 \\
 T10  & 4.8    & 0.055   &  0.07   & 0.055 \\
 B05  & 4.7    & 0.072   &  0.010  & 0.055 \\
 \hline
\end{tabular}

\medskip
Here $\rho_{M}$ is the core mass density in  $M_\odot\,{\rm pc}^{-3}$,
$f_{\rm b,c}$ is the binary fraction in the core, $f_{0.5}$ is the  binary fraction
for non-degenerate stars more massive than $0.5\,M_\odot$, and
$f_{\rm wd}$ is the binary fraction among white dwarfs.
Values for all quantities are given at 14 Gyr.
\end{table}

\subsection{Main Reference Model}\label{sec:mainrefmodel}

First we present our results for a typical dense cluster, represented  by Model~1.
With 100\% binaries initially, the final core binary fraction
(at 14~Gyr), $f_{\rm b,c}$, is only 9.5\%.
This is strikingly low, given that the cluster started with {\em all\/}  stars
in binaries, and that binaries should concentrate more into the core  through mass segregation,
but it is expected from our estimates in \S~4.
Decreasing the initial binary fraction, $f_{\rm b,0}$, to a more  reasonable
(but still large) 50\% reduces $f_{\rm b,c}$ further to 7\%, as shown  in Model~B05
(see Table~\ref{table2}).
The dependence of $f_{\rm b,c}$ on $f_{\rm b,0}$ is not linear. This is
mainly due to mass segregation: decreasing $f_{\rm b,0}$ also increases  the
ratio of mean binary mass to mean stellar mass in the cluster, thereby  resulting
in a higher concentration of binaries in the core.

The majority (about 75\%) of destroyed binaries were disrupted by close  dynamical encounters
(or, rarely, following a supernova explosion). Note that some binaries  that are initially hard
eventually become soft after undergoing significant mass loss due to  stellar evolution.
About 20\% of the destroyed binaries experienced mergers, typically  after significant hardening through interactions.
A few percent of the binaries lost were actually not destroyed but  instead were ejected
from the cluster as a result of recoil in strong encounters.
Tidal capture did not play a significant role:
the total number of tidal-capture binaries formed during the cluster  lifetime is less than 1\% of the final number of binaries in the core.
The total core mass during most of the cluster evolution (the last  $\sim 10\,t_{\rm rh}$) did not vary by more than a factor of two.

While the final core binary fraction is extremely low, the overall  cluster
binary fraction, which takes into account all halo binaries, remains  high.
However, the halo binaries consist mainly of very
low-mass systems: the average primary mass among binaries
remaining outside the core at 14~Gyr is $0.2\,M_\odot$, with the  average companion
mass around $0.1\,M_\odot$. These binaries would be extremely faint and  hard to
detect observationally.

\begin{figure}
\includegraphics[width=84mm]{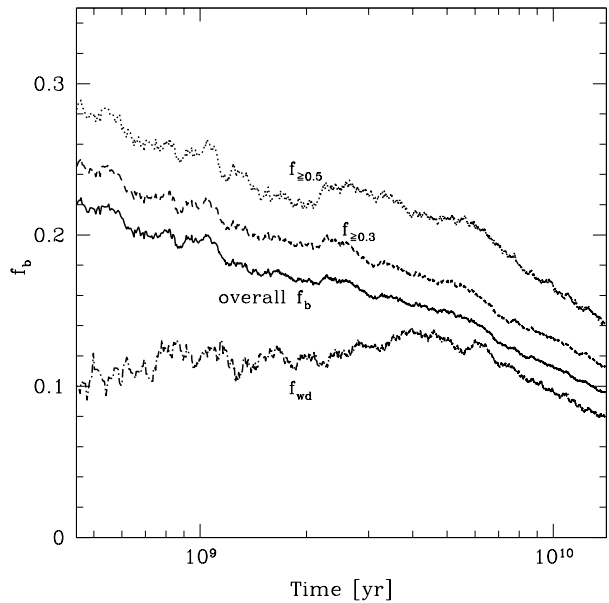}
\caption{Evolution of the core binary fraction in Model~1.
We show separately the binary fractions for all objects, for
non-degenerate stars with mass $\ge 0.5\,M_\odot$,
for non-degenerate stars with mass $\le 0.3\,M_\odot$, and for white  dwarfs.}
\label{bf_mb}
\end{figure}

We have also examined different stellar subpopulations in the cluster:
(i) the subpopulation of  non-degenerate objects with mass (for a  single star
or for the primary in a binary) $\ge 0.5\,M_\odot$ and (ii)  the  subpopulation
of all white dwarfs, single or in binaries.
The binaries in group (i) may be easier to detect, e.g., from  broadening of
the main sequence in a colour-magnitude diagram \citep{1997ApJ...474..701R}.
Binaries in group (i) are harder than less massive average binaries,
so fewer of them are destroyed.
On the other hand, stellar evolution plays a more significant role
in the destruction of white-dwarf binaries, which were initially more  massive
and harder (Fig.~\ref{bf_mb}).
Therefore the binary destruction rate in group (ii) is much higher,
enhanced both by stellar evolution
(mass loss and mass transfer at more advanced evolutionary stages,
and supernovae in binaries), and by dynamical interactions
(larger cross-section for encounters).
Note also that the overall $f_{\rm b,c}$ is decreased partially
through a lower binary fraction for degenerate objects.

\begin{figure}
\includegraphics[width=84mm]{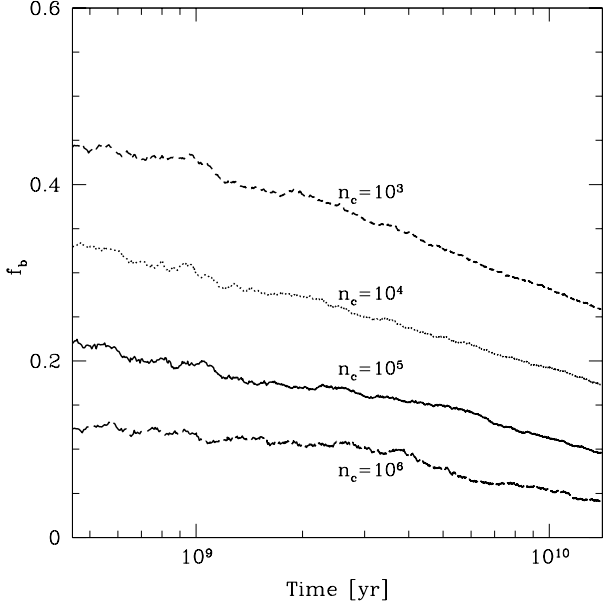}
\caption{Evolution of the core binary fraction in Model~1 compared
to models with different core number densities
(D3, D4 and D6).}
\label{bf_dens}
\end{figure}

\subsection{Different Densities and Relaxation Times}

Let us now compare results for different central densities, in Models  1, D3, D4 and D6. The evolution
of $f_{\rm b,c}$ for these models is shown in Figure~\ref{bf_dens}.
As expected, the core binary fraction decreases as $n_c$ increases.
The dependence is steeper at high densities, where dynamical interaction
rates play a more
dominant role compared to stellar evolution.

With respect to the half-mass relaxation time (Models T8 and T10),
we find that, surprisingly, the model with shorter relaxation time
has a higher core binary fraction.
There are two competing mechanisms that play a role here: mass  segregation,
which brings binaries into the core, and dynamical interactions, which  destroy
binaries in the core.
A shorter segregation time increases the rate at which binaries enter  the core but also
allows less massive binaries  to interact.
Therefore, the average mass of a binary in the  core and the effective
mass density in the core is smaller. As a result, fewer binaries are  destroyed.

\subsection{Binary Period Distribution}

\begin{figure}
\includegraphics[width=84mm]{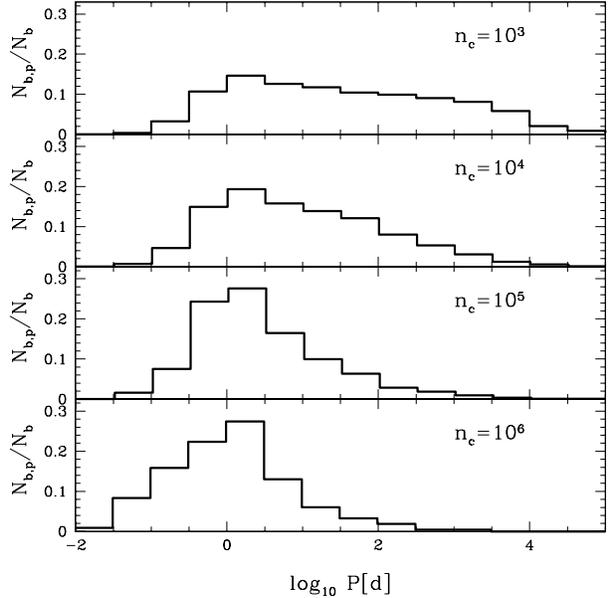}
\caption{The binary period distributions in models with different  densities (from the top: D3, D4, Model~1 and D6).
N$_{\rm b}$ is the total number of binaries and N$_{\rm b,p}$
is the number of binaries in each period bin.}
\label{periods_all}
\end{figure}

\begin{figure}
\includegraphics[width=84mm]{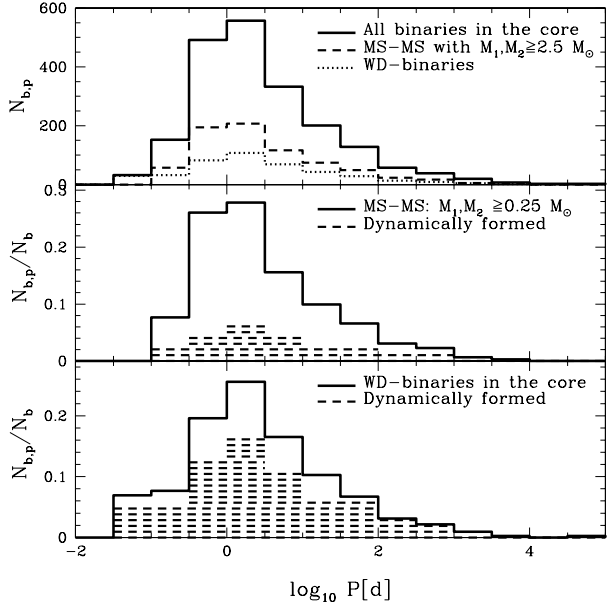}
\caption{Period distributions for different binaries in Model~1 (at 14  Gyr).
The middle plot shows the period
distribution of binaries containing two MS stars with
masses greater than $0.25\,M_\odot$, bottom plot shows
 the period distribution of binaries with at least one WD;
shaded area shows the fraction of dynamically formed binaries.
N$_{\rm b}$ is the total number of binaries and N$_{\rm b,p}$
is the number of binaries in each period bin.}
\label{p_complete}
\end{figure}

Through dynamical interactions, we would expect that the initial
period distribution of binaries
should be depleted above the boundary between hard and soft binaries.
Stellar evolution should deplete a fraction of hard binaries,
especially at very short periods, and
dynamical encounters should further deplete some of the wider hard  binaries.
Indeed, for lower-density clusters, we find that the distribution  remains
much flatter in $\log P$ (see Fig.~\ref{periods_all}),
with more and more of the wider hard binaries disappearing as the
density increases.

This period distribution can be used to better extrapolate observed
binary fractions, which are usually limited to rather narrow ranges
of masses and periods. In particular, it is clear that the usual
assumption of a flat distribution in $\log P$ for hard binaries
{\em at present\/} in a cluster core \citep[e.g.,][]{BinFreq_47Tuc_01} can be very misleading.
This will be done in \S~5.5 for our models of several real clusters.

\subsection{Comparison with Observations}

\begin{table}
\caption{Results for Models of Specific Clusters}
\label{table3}
\begin{tabular}{@{}l l l l l l }
\hline
{Model}    & Age &    {$\log \rho_{ M}$}
& {$f_{b,c}$}
 & {$f_{0.5}$}         & {$f_{\rm wd}$} \\
\hline
M12    & 12  & 3.5 & 0.170  & 0.26 & 0.130 \\
M4     & 13 & 4.4 & 0.115 & 0.175 & 0.10 \\
47~Tuc  & 11 & 5.1 & 0.080 & 0.125  & 0.075 \\
NGC~6388& 12  & 5.7 & 0.06 & 0.08 & 0.045 \\
\hline
\end{tabular}

\medskip
With the same notations as before, the columns give:
the name of the cluster;
the cluster age in Gyr;
the central mass density in $M_\odot\,{\rm pc}^{-3}$ in simulations at the given age,
the core binary fraction, the binary fraction
for non-degenerate objects more massive than $0.5\,M_\odot$, and
the binary fraction among white dwarfs.
\end{table}

\begin{figure}
\includegraphics[width=84mm]{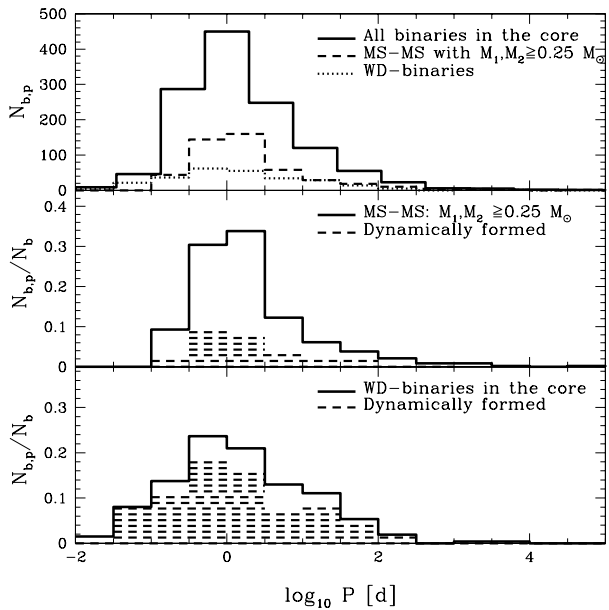}
\caption{The binary period distribution in our 47~Tuc Model, at 11 Gyr.
The middle plot shows the period
distribution of binaries containing two main-sequence stars with
masses greater than $0.25\,M_\odot$; the bottom plot shows
 the period distribution of binaries with at least one white-dwarf  component
(the shaded area shows the fraction of dynamically formed binaries).
As before N$_{\rm b}$ is the total number of binaries and N$_{\rm b,p}$
is the number of binaries in each period bin.}
\end{figure}

We performed several simulations with parameters that attempt to match  those of specific globular
clusters in the Galaxy (Tables~\ref{table1} and~\ref{table3}).
For these models, the total core masses we compute at present match  well those of
King models fitted to the observed cluster parameters, and
the corresponding core radii of our models (from total core mass and  density)
are all $\sim 0.5\,$pc, matching the observed values.
In all cases the initial binary fraction is assumed to be 100\%,
so our results for final core binary fractions represent upper limits.
As in all reference models, we predict low values for $f_{\rm b,c}$
in all globular clusters, with smaller values in higher-density cores.

We compare our derived binary fraction for M4 with the results from \citet{1996AJ....112..565C}.
Using a Monte Carlo modelling of the binary population, they found that
the best fit to the observed binaries (in the period range from 2~days
to 3~years) is $f_{\rm b,c}\simeq15\%$.
This is in good agreement with our overall predicted $f_{\rm  b,c}=11.5\%$, and
also with our prediction for heavier main-sequence binaries,  $f_{0.5}=17.5\%$.

In the 47~Tuc model (for which we increased the total number of stars
initially to a more realistic $N=5\times 10^5$),
$f_{\rm b,c}$ is only 8\% at an age of 11 Gyr and about 7\% at 14 Gyr
(we provide this value at different ages given the uncertainty in 
observationally determined values for the ages of 47~Tuc; see, e.g.,
Schiavon et al. 2002; Zoccali at al. 2001;
also note that $f_{\rm b,c}$ does not change much
over the last several Gyr).
At first glance, this may seem to conflict with observations.
In particular, \cite{BinFreq_47Tuc_01} derive an overall binary  fraction for the core
of 47~Tuc of about 13\%, from observations of eclipsing binaries with  periods
in the range $P\simeq 4$--$16\,$d.
This estimate was based on an extrapolation
assuming a period distribution flat in $\log P$ from about $2.5\,$d to  $50\,$yr
(soft primordial binaries with $P> 50\,$yr are assumed destroyed 
and  short-period primordial binaries with $P\le 4\,$d are assumed to evolve toward much  shorter periods
through angular momentum loss by magnetic braking).
In Figure~2 we show the final period distribution of core binaries in  our simulation.
Note that the period range of eclipsing binaries is near the peak
of the distribution, while for longer periods the number of binaries
drops rapidly.
In particular, if we concentrate on the binaries consistent with the  observed set,
with primary masses $M_1 > 0.25 M_\odot$ and $q > 0.3$, we find that
the number of systems with periods in the range $16\,$d to $50\,$yr
is about 6 times smaller than would be predicted by adopting a  distribution
flat in $\log P$.
For the whole period range from $2.5\,$d to $50\,$yr we have 2.2 times
fewer binary systems compared to the flat distribution.
If we take into account this depletion of wider binaries when modelling
the number of observed eclipsing binaries in 47~Tuc, we are led to
revise the observed core binary fraction from \citet{BinFreq_47Tuc_01}
to  $6\%\pm2$, which is much closer to our theoretical prediction.

We performed three additional simulations for 47~Tuc,
with $f_{\rm b,0}=0.25$, 0.5, and 0.75.
The corresponding core binary fraction extrapolated from observations  (corrected as
above) does not vary much among these different models.
Its maximum value is obtained for $f_{\rm b,0}=0.5$, which gives
a core binary fraction of about 8\%, and in this case the total number  of binary systems
is 1.6 times less than with an assumed flat distribution.

An alternative estimate of the binary fraction in the core of 47~Tuc is  based on observations
of BY~Dra stars \citep{BinFreq_47Tuc_01}.  Their estimated core binary  fraction,
which can be considered a {\em lower limit\/}, is approximately 0.8\%,
18 times lower than the estimate based on eclipsing binaries.
This estimate was based on 31 BY~Dra binaries 
(observed in the period range 0.4--10 d) and 5 eclipsing binaries (period range 4--16 d).
We analysed the core binary population in our model in order to identify BY~Dra
systems and eclipsing binaries, and considering the ratio of the two.
We adopted the standard definition for a BY~Dra binary: primary mass in  the range
$0.3$--$0.7\,M_\odot$ \citep[see, e.g.,][]{BoppFekel_BYDra_77}
and period in the range $0.4$--$10\,$d (as for the observed sample in  47~Tuc).
For eclipsing binaries we adopted the observed period range 4--16 d with each star
$\ge 0.25 M_\odot$.
The ratio between the number of BY~Dra systems and the number of  eclipsing binaries is
found to be 5.9, 6.7, 7.2 and 3.5 for models with $f_{\rm  b,0}=1.0,0.75,0.5$ and 0.25, respectively.
Therefore a large initial binary fraction $\ga 75\%$ is most likely.

Of the quantities that we explicitly assume in our cluster model
to be constant, the central (core) density is the one with the largest
dynamic range in models that provide for dynamical evolution.  Hence it
is also the quantity that is most likely to significantly affect our final
results.
To test the sensitivity of our results to a changing central density,
we have run our model of 47~Tuc with a central density assumed to
increase by a factor of 10 from $t=0$ to the present. Specifically,
we still match the currently observed core density while ramping up the
value either exponentially or linearly in time,
starting with a value 10 times smaller than at present. This could
represent qualitatively the gradual core contraction observed in some
$N$-body models of star clusters with binaries 
\citep[see][and, for a steeper core contraction, see Giersz \& Spurzem 2003]{1993gcgc.work..701A}.
These models predict a present-day core binary fraction that is only  slightly higher, by about 1--2\%.
To increase the binary fraction more significantly,
the time-averaged core density in the cluster would have had to be many
orders of magnitude lower than what is observed today. There is no  reasonable
 dynamical history that would produce such an unusual result.
In contrast, recent $N$-body simulations show that 
the presence of just 10\% hard primordial binaries leads to core radius {\it expansion} 
and therefore the core density might be higher in the past (Wilkinson  et al.\ 2003).
Though we did not run another 47~Tuc model with a central density 
decreasing with time, we can predict that the present-day binary fraction would then be smaller by about 1--2\%.

\section{Conclusions}

We have considered in detail processes of binary destruction (and formation) in dense stellar systems.
In particular, we have shown that the present binary fraction in cluster cores should be relatively small
($\la 10\%$).  This is caused not only by  dynamical encounters, but also binary stellar evolution, which is
the dominant mechanism for the destruction of hard  binaries.
We also find that the destruction rate due to stellar evolution is enhanced significantly
by dynamical hardening of binaries.

We have shown that values of binary fractions for stars in different
mass ranges and at different evolutionary stages can differ  significantly.
The fraction of dynamically formed binaries is higher when one
considers stars at more advanced evolutionary stages.  The binary period
distribution evolves from flat in $\log P$ for loose clusters toward a
sharply peaked distribution in denser clusters, even if all clusters  have
identical velocity dispersions and therefore identical hard-soft binary boundaries.
This implies that a flat period distribution should not be assumed
when deriving overall
binary fractions by extrapolation from the distribution of observed
binaries in a narrow period range (e.g., eclipsing binaries).

We considered several models that attempted to match observed globular  clusters.
For those with good available data on binaries (M4 and 47~Tuc),
we found our predicted binary fractions to be in good agreement with  observations
once we take into account the correct binary
period distribution. The main conclusion we derive from these  calculations
and our semi-analytic estimates is that
the currently observed binary fractions in cluster cores suggest very
high initial (primordial) binary fractions---close to 100\%.

In addition to their implications for the interpretation of observed  binary fractions
in cluster cores, our results have important consequences for the
theoretical modeling of globular clusters using $N$-body simulations.
Indeed, it is clear that ``realistic'' dynamical
simulations of globular cluster evolution should include large  populations of
primordial binaries, with initial binary fractions in the range $\sim  50\%$--$100\%$
\citep[similar to what is usually assumed for the field; see, e.g.,][]{Multiple_91}. This poses a
particular challenge for direct $N$-body simulations, where the  treatment of even
relatively small numbers of binaries can add enormous computational  costs.  For this
reason, current direct $N$-body simulations of star clusters with large  initial
binary fractions typically have $N$ too small to be considered  representative of
globular clusters (see, e.g., Portegies Zwart et al.\ 2003; Wilkinson  et al.\ 2003).
Other methods, such as Monte Carlo simulations, do not suffer from the  same
limitations, and routinely simulate clusters with reasonably large $N$
($\sim 10^5 - 10^6$) and binary fractions ($\sim 30\%$), but have not  yet
included advanced treatments of binary star evolution
\citep[see, e.g., ][and references therein]{2003ApJ...593..772F}.
          
\section*{Acknowledgments}
We thank D.\ Heggie and  J.\ Hurley for
very helpful discussions. This work was supported by NSF Grant AST-0206276,
NASA ATP Grant NAG5-12044, and a Chandra Theory grant.

{

}

\appendix

\section{Role of Three-Body Binary Formation}

Consider the rate of three-body binary formation (via the close approach of
three single stars) in a dense cluster core.  We denote by $\Gamma(E_{\rm b})$ the rate per 
star of the formation of a binary with binding energy $E_{\rm b}$.
First, we consider $\Gamma (b\le b_{\rm max})$ -- the rate (per star) at which three objects
come together within a region of size $b_{\rm max}$.
The probability that a third object will be in the vicinity $b$ of two  other interacting objects is
the product of the probability of the first two bodies meeting and the
probability that during this time
a third object will be in the same vicinity. The rate of two-body
encounters for masses $m_1$ and $m_2$, with number density $n_2$ is

\begin{equation}
\Gamma_2 (b\le b_{\rm max}) = \pi b_{\rm max}^2 \left (1 + {\frac  {v^2_{p}} {<v_{12}>^2} } \right ) n_2 v_{12} .
\end{equation}
Here 
\begin{equation}
v_{p}^2 = {\frac {2 G (m_1+m_2)} {b_{\rm max}}}
\end{equation}
is the velocity at closest approach and
\begin{equation}
<v_{12}>^2\simeq  (\sigma_1^2 + \sigma_2^2) =  \sigma^2 <m> {\frac  {m_1 + m_2} {m_1 m_2}},
\end{equation}
where $\sigma$ is the three-dimensional velocity dispersion.

We define the minimum hardness for the binary to be formed as
\begin{equation}
\eta_{\rm min} = {\frac {G m_1 m_2} {b_{\rm max} <m> \sigma^2}}.
\end{equation}
Then
\begin{equation}
\Gamma_2 (b\le b_{\rm max}) = \pi b_{\rm max}^2 \left (1 + 2\eta  \right ) n_2 v_{12}.
\end{equation}

The second object spends in the vicinity $b$ of the first object a  time
$\Delta t \simeq {\frac {2 b } {v_{p}}}$. The probability that a third  object
will be within the same vicinity is then
\begin{equation}
p_3 \simeq  n_3 (b_{\rm max}^2  \Delta t v_3 + b_{\rm max}^3) =
n_3 b_{\rm max}^3 \left ( 1 + 2 {\frac {v_3} {v_p} }\right).
\end{equation}
Here $v_3$ is the relative velocity of the third object with respect to
the centre of mass of the first two.
Effectively, it reflects
velocity at which the population at the vicinity $b$ is increasing.
We do not consider here the population decrease, assuming that
if the object was in the neighborhood, a three-body encounter has  happened.

The final result then takes the form given in (\cite{Binney_GC_87}, \S  8), but
with an additional mass- and hardness-dependent factor $f$,
\begin{equation}
\label{g3}
\Gamma_3 (\eta > \eta_{\rm min}) = {\frac { n_c^2 G^5 <m>^5}  {\sigma^9}}  f(m_1,m_2, m_3,\eta)
\end{equation}
\begin{equation}
\begin{array} {ll}
f(m_1,m_2, m_3,\eta) =&   \\
& \pi  {\frac {n_2 n_3} {n_c^2}}  {\frac {m_1^5} {<m>^5}}  {\frac  {m_2^5} {<m>^5}}   \eta^{-5} \left (1 + 2 \eta \right )  \\
& \left ( 1 + {\frac {v_3} {\sigma}} \eta^{-0.5} \sqrt{2 \frac {m_1 m_2  } {(m_1+m_2)<m> } }\right) {\frac {v_{12}} {\sigma}} .
\end{array}
\end{equation}
It is clear that the rate is highly dependent on the masses of the  participating
stars and decreases steeply with increasing hardness of the binary that  is formed.

If all encounters resulted in the formation of a binary with hardness  $\eta$,
then the rate of binary formation
would be completely described by equation~\ref{g3}.

Using the equation~(\ref{g3}), the time-scale for three-body binary formation
can also be written as
\begin{equation}
\begin{array}{ll}
T_{3b} = &
2.08 \times 10^{16} [yr] \ \left ( {\frac {10^5} {n_c}}\right ) ^2   \left ( {\frac {M_\odot} {<m>}}\right ) ^{5}
\left ( {\frac {\sigma} {10\ {\rm km/s}}}\right ) ^{9} \\
& {\frac {n_c} {n_2}}{\frac {n_c} {n_3}}
\left ( {\frac {<m>} {m_1}}\right ) ^{5} \left ( {\frac {<m>}  {m_2}}\right ) ^{5} {\frac {\sigma} {v_{12}}} \\
& \eta^5 (1+2\eta)^{-1} \left ( 1 + {\frac {v_3} {\sigma}} \eta^{-0.5}  \sqrt{2 \frac {m_1 m_2 } {(m_1+m_2)<m> } }\right)^{-1}
\end{array}
\end{equation}

Even in the core of a large, very dense cluster,
 with density $\sim10^5\,{\rm pc}^{-3}$ and
containing $\sim10^5$ stars,
no binaries with $\eta > 1$ will be formed from $\sim 1\,M_\odot$ stars  in
a Hubble time. 
It has also been shown that many three-body binary formation events will
lead to physical collisions for stars as small as white dwarfs \citep{1996IAUS..174..263C}.
We therefore neglect all three-body interactions in our cluster simulations.

\label{lastpage}


\begin{thebibliography}{66}
\expandafter\ifx\csname natexlab\endcsname\relax\def\natexlab#1{#1}\fi

\bibitem[Aarseth(2001)]{2001NewA....6..277A} Aarseth, S.~J.\ 2001, New 
Astronomy, 6, 277 

\bibitem[Aarseth \& Heggie(1993)]{1993gcgc.work..701A} Aarseth, S.~J.~\& 
Heggie, D.~C.\ 1993, ASP Conf.~Ser.~ 48: The Globular Cluster-Galaxy 
Connection, 701 

\bibitem[{{Albrow} {et~al.}(2001){Albrow}, {Gilliland}, {Brown},  {Edmonds},
  {Guhathakurta}, \& {Sarajedini}}]{BinFreq_47Tuc_01}
{Albrow}, M.~D., {Gilliland}, R.~L., {Brown}, T.~M., {Edmonds}, P.~D.,
  {Guhathakurta}, P., \& {Sarajedini}, A. 2001, ApJ, 559, 1060

\bibitem[{{Bacon} {et~al.}(1996){Bacon}, {Sigurdsson}, \&
  {Davies}}]{1996MNRAS.281..830B}
{Bacon}, D., {Sigurdsson}, S., \& {Davies}, M.~B. 1996, MNRAS, 281, 830

\bibitem[{{Belczynski} {et~al.}(2002){Belczynski}, {Kalogera}, \&  {Bulik}}]{Chris_02}
{Belczynski}, K., {Kalogera}, V., \& {Bulik}, T. 2002, ApJ, 572, 407

\bibitem[{{Belczynski} {et~al.}(2005){Belczynski}, {Kalogera}, {Rasio},  \& {Taam}}]{Chris_04}
{Belczynski}, K., {Kalogera}, V., {Rasio},F., \& {Taam}, R. 2005, {in  prep}

\bibitem[Bellazzini et al.(2002a)]{2002AJ....123.1509B} Bellazzini, M.,  Fusi
Pecci, F., Messineo, M., Monaco, L., \& Rood, R.~T.\ 2002a, AJ, 123,  1509

\bibitem[{{Bellazzini} {et~al.}(2002b){Bellazzini}, {Fusi Pecci},  {Montegriffo},
  {Messineo}, {Monaco}, \& {Rood}}]{2002AJ....123.2541B}
{Bellazzini}, M., {Fusi Pecci}, F., {Montegriffo}, P., {Messineo}, M.,
  {Monaco}, L., \& {Rood}, R.~T. 2002b, Astron. J., 123, 2541

\bibitem[Benacquista(2002)]{2002LRR.....5....2B} Benacquista, M.\ 2002,
Living Reviews in Relativity, 5, 2

\bibitem[Bethe \& Brown(1998)]{1998ApJ...506..780B} Bethe, H.~A.~\&  Brown,
G.~E.\ 1998, ApJ, 506, 780


\bibitem[{{Binney} \& {Tremaine}(1987)}]{Binney_GC_87}
{Binney}, J., \& {Tremaine}, S. 1987, {Galactic dynamics} (Princeton,  NJ,
  Princeton University Press, 1987, 747 p.)

\bibitem[{{Bopp} \& {Fekel}(1977)}]{BoppFekel_BYDra_77}
{Bopp}, B.~W., \& {Fekel}, F. 1977, Astron. J., 82, 490

\bibitem[Brandner et al.(1996)]{1996A&A...307..121B} Brandner, W.,  Alcala,
J.~M., Kunkel, M., Moneti, A., \& Zinnecker, H.\ 1996, A\&A, 307, 121

\bibitem[Chernoff \& Huang(1996)]{1996IAUS..174..263C} Chernoff, D.~F.~\& 
Huang, X.\ 1996, IAU Symp.~174: Dynamical Evolution of Star Clusters: 
Confrontation of Theory and Observations, 174, 263

\bibitem[{{Cool} \& {Bolton}(2002)}]{CoolBolton_NGC6397_02}
{Cool}, A.~M., \& {Bolton}, A.~S. 2002, in ASP Conf. Ser. 263: Stellar
  Collisions, Mergers and their Consequences, 163

\bibitem[Cote \& Fischer(1996)]{1996AJ....112..565C} Cote, P.~\&  Fischer,
P.\ 1996, Astron. J., 112, 565

\bibitem[{{Cote} {et~al.}(1996){Cote}, {Pryor}, {McClure}, {Fletcher},  \&
  {Hesser}}]{Cote_M22_96}
{Cote}, P., {Pryor}, C., {McClure}, R.~D., {Fletcher}, J.~M., \&  {Hesser},
  J.~E. 1996, Astron. J., 112, 574

\bibitem[Davies(1995)]{1995MNRAS.276..887D} Davies, M.~B.\ 1995, MNRAS, 
276, 887 

\bibitem[Davies \& Benz(1995)]{1995MNRAS.276..876D} Davies, M.~B.~\& Benz, 
W.\ 1995, MNRAS, 276, 876 

\bibitem[Davies(1997)]{1997MNRAS.288..117D}
Davies, M.~B.\ 1997, MNRAS, 288, 117

\bibitem[Di Stefano \& Rappaport(1994)]{1994ApJ...437..733D} Di Stefano,
R.~\& Rappaport, S.\ 1994, ApJ, 437, 733

\bibitem[{{Dubath} {et~al.}(1997){Dubath}, {Meylan}, \&
  {Mayor}}]{Dubath_Disp_97}
{Dubath}, P., {Meylan}, G., \& {Mayor}, M. 1997, A\&A, 324, 505

\bibitem[{{Duquennoy} \& {Mayor}(1991)}]{Multiple_91}
{Duquennoy}, A., \& {Mayor}, M. 1991, A\&A, 248, 485


\bibitem[Freire et al.(2001)]{2001MNRAS.326..901F} Freire, P.~C.,  Camilo,
F., Lorimer, D.~R., Lyne, A.~G., Manchester, R.~N., \& D'Amico, N.\  2001,
MNRAS, 326, 901

\bibitem[{{Fregeau} {et~al.}(2004){Fregeau}, {Cheung}, {Portegies
  Zwart}, \& {Rasio}}]{Fregeau_FB2_04}
Fregeau, J.~M., Cheung,  P., Portegies Zwart, S.~F., \& Rasio, F.~A.\ 2004, MNARS, 352, 1 


\bibitem[{{Fregeau} {et~al.}(2003){Fregeau}, {G{\" u}rkan},
  {Joshi}, \& {Rasio}}]{2003ApJ...593..772F}
{Fregeau}, J.~M., {G{\" u}rkan}, M.~A., {Joshi}, K.~J., \& {Rasio},  F.~A.
  2003, ApJ, 593, 772

\bibitem[{{Fregeau} {et~al.}(2002){Fregeau}, {Joshi}, {Portegies  Zwart}, \&
  {Rasio}}]{Fregeau_MS_02}
{Fregeau}, J.~M., {Joshi}, K.~J., {Portegies Zwart}, S.~F., \& {Rasio},  F.~A.
  2002, ApJ, 570, 171

\bibitem[{{Fryer} \& {Kalogera}(2001)}]{FryerKalogera_BH_01}
{Fryer}, C.~L., \& {Kalogera}, V. 2001, ApJ, 554, 548

\bibitem[{{Gao} {et~al.}(1991){Gao}, {Goodman}, {Cohn}, \&
  {Murphy}}]{Gao_FP_91}
{Gao}, B., {Goodman}, J., {Cohn}, H., \& {Murphy}, B. 1991, ApJ, 370,  567

\bibitem[Giersz \& Spurzem(2003)]{2003MNRAS.343..781G} Giersz, M.~\& 
Spurzem, R.\ 2003, MNRAS, 343, 781 

\bibitem[{{Goodman} \& {Hut}(1989)}]{GoodmanHut_89}
{Goodman}, J., \& {Hut}, P. 1989, Nat, 339, 40

\bibitem[Goldberg, Mazeh, \& Latham(2003)]{2003ApJ...591..397G}  Goldberg,
D., Mazeh, T., \& Latham, D.~W.\ 2003, ApJ, 591, 397

\bibitem[Gratton et al.(2003)]{2003A&A...408..529G} Gratton, R.~G.,
Bragaglia, A., Carretta, E., Clementini, G., Desidera, S., Grundahl,  F., \&
Lucatello, S.\ 2003, A\&A, 408, 529


\bibitem[G{\" u}rkan, Freitag, \& Rasio(2004)]{2004ApJ...604..632G} G{\"
u}rkan, M.~A., Freitag, M., \& Rasio, F.~A.\ 2004, ApJ, 604, 632

\bibitem[Halbwachs, Mayor, Udry, \& Arenou(2003)]{2003A&A...397..159H}
Halbwachs, J.~L., Mayor, M., Udry, S., \& Arenou, F.\ 2003, A\&A, 397,  159

\bibitem[{{Harris}(1996)}]{Harris_GCcatalog_96}
{Harris}, W.~E. 1996, Astron. J., 112, 1487

\bibitem[{{Heggie}(1974)}]{1974CeMec..10..217H}
{Heggie}, D.~C. 1974, Celestial Mechanics, 10, 217

\bibitem[Hills(1984)]{1984AJ.....89.1811H} Hills, J.~G.\ 1984, Astron. J., 89, 1811

\bibitem[{{Hills}(1990)}]{Hills_90}
{Hills}, J.~G. 1990, Astron. J., 99, 979



\bibitem[Hurley \& Shara(2003)]{2003ApJ...589..179H} Hurley, J.~R.~\&
Shara, M.~M.\ 2003, ApJ, 589, 179

\bibitem[{{Hurley} {et~al.}(2000){Hurley}, {Pols}, \&
  {Tout}}]{Hurley_Single_00}
{Hurley}, J.~R., {Pols}, O.~R., \& {Tout}, C.~A. 2000, MNRAS, 315, 543

\bibitem[{{Hurley} {et~al.}(2002){Hurley}, {Tout}, \&
  {Pols}}]{Hurley_Binary_02}
{Hurley}, J.~R., {Tout}, C.~A., \& {Pols}, O.~R. 2002, MNRAS, 329, 897

\bibitem[Hut, McMillan, \& Romani(1992)]{hut_1992} Hut, P.,
McMillan, S., \& Romani, R.~W.\ 1992, ApJ, 389, 527

\bibitem[Iben \& Livio(1993)]{1993PASP..105.1373I} Iben, I.~J.~\& Livio, 
M.\ 1993, PASP, 105, 1373 

\bibitem[Ivanova \& Rasio(2004a)]{Aspen_2004} Ivanova, N.~\& Rasio, 
F.\ 2004a, to appear in "Binary Radio Pulsars," ASP Conf. Series, ed. F.A. Rasio \& I.H. Stairs,
astro-ph/0405382

\bibitem[Ivanova \& Rasio(2004b)]{2004RMxAC..20...67I} Ivanova, N.~\& Rasio, 
F.\ 2004b, Revista Mexicana de Astronomia y Astrofisica Conference Series, 
20, 67 

\bibitem[Ivanova \& Taam(2003)]{2003ApJ...599..516I} Ivanova, N.~\&  Taam,
R.~E.\ 2003, ApJ, 599, 516

\bibitem[Joshi, Rasio, \& Portegies Zwart(2000)]{2000ApJ...540..969J}
Joshi, K.~J., Rasio, F.~A., \& Portegies Zwart, S.\ 2000, ApJ, 540, 969

\bibitem[Joshi, Nave, \& Rasio(2001)]{2001ApJ...550..691J} Joshi, K.~J.,
Nave, C.~P., \& Rasio, F.~A.\ 2001, ApJ, 550, 691

\bibitem[{{King}(1965)}]{King_65}
{King}, I.~R. 1965, Astron. J., 70, 376

\bibitem[K{\" o}hler, Leinert, \& Zinnecker(2001)]{2001AGM....18..P24K}
K{\" o}hler, R., Leinert, C., \& Zinnecker, H.\ 2001, Astronomische
Gesellschaft Meeting Abstracts, 18, 24

\bibitem[{{Kroupa}(2002)}]{Kroupa_IMF_02}
{Kroupa}, P. 2002, Science, 295, 82

\bibitem[McMillan, McDermott, \& Taam(1987)]{1987ApJ...318..261M} McMillan, 
S.~L.~W., McDermott, P.~N., \& Taam, R.~E.\ 1987, ApJ, 318, 261 

\bibitem[Melo(2003)]{2003A&A...410..269M} Melo, C.~H.~F.\ 2003, A\&A,  410, 269

\bibitem[{{Mikkola}(1983)}]{1983MNRAS.203.1107M}
{Mikkola}, S. 1983, MNRAS, 203, 1107

\bibitem[{{Mikkola}(1985)}]{1985MNRAS.215..171M}
{Mikkola}, S. 1985, MNRAS, 215, 171


\bibitem[Portegies Zwart et al. (2001)]{2001MNRAS.321..199P} Portegies Zwart, S.F., McMillan,
S.L.W., Hut, P., \& Makino, J.\ 2001, MNRAS, 321, 199

\bibitem[Portegies Zwart, Hut, McMillan, \& Verbunt(1997a)]
{ecology_i} Portegies Zwart, S.~F., Hut, P.,
McMillan, S.~L.~W., \& Verbunt, F.\ 1997a, A\&A, 328, 130


\bibitem[Portegies Zwart, Hut, McMillan, \& Verbunt(1997b)]
{ecology_ii} Portegies Zwart, S.~F., Hut, P.,
McMillan, S.~L.~W., \& Verbunt, F.\ 1997b, A\&A, 328, 143

\bibitem[{{Portegies Zwart et~al.}(2003){Portegies Zwart}, {Hut},  {McMillan},
  \& {Makino}}]{ecology_v}
Portegies 
Zwart, S.~F., Hut, P., McMillan, S.~L.~W., \& Makino, J.\ 2004, MNRAS, 
351, 473 

\bibitem[{{Portegies Zwart} \& {Meinen}(1993)}]{Zwart_TC_93}
{Portegies Zwart}, S.~F., \& {Meinen}, A.~T. 1993, A\&A, 280, 174

\bibitem[{{Pryor} \& {Meylan}(1993)}]{Pryor_GCdisp_93}
{Pryor}, C., \& {Meylan}, G. 1993, in ASP Conf. Ser. 50: Structure and  Dynamics
  of Globular Clusters, 357

\bibitem[Rasio, Pfahl, \& Rappaport(2000)]{2000ApJ...532L..47R} Rasio,
F.~A., Pfahl, E.~D., \& Rappaport, S.\ 2000, ApJL, 532, L47

\bibitem[{{Rubenstein} \& {Bailyn}(1997)}]{1997ApJ...474..701R}
{Rubenstein}, E.~P., \& {Bailyn}, C.~D. 1997, ApJ, 474, 701


\bibitem[Schiavon, Faber, Rose, \& Castilho(2002)]{2002ApJ...580..873S} 
Schiavon, R.~P., Faber, S.~M., Rose, J.~A., \& Castilho, B.~V.\ 2002, ApJ, 
580, 873 

\bibitem[Sigurdsson \& Phinney(1995)]{1995ApJS...99..609S} Sigurdsson,
S.~\& Phinney, E.~S.\ 1995, ApJS, 99, 609

\bibitem[{{Sills et~al.}(2003)}]{Modest2}
{Sills}, A. {et al.} 2003, New Astronomy, 8, 605

\bibitem[Shara \& Hurley(2002)]{2002ApJ...571..830S} Shara, M.~M.~\&
Hurley, J.~R.\ 2002, ApJ, 571, 830

\bibitem[Smith \& Bonnell(2001)]{2001MNRAS.322L...1S} Smith, K.~W.~\&
Bonnell, I.~A.\ 2001, MNRAS, 322, L1

\bibitem[Tokovinin(1997)]{1997AstL...23..727T} Tokovinin, A.~A.\ 1997,
Astronomy Letters, 23, 727

\bibitem[Verbunt \& Zwaan(1981)]{1981A&A...100L...7V} Verbunt, F.~\& Zwaan, 
C.\ 1981, A\&A, 100, L7 

\bibitem[{{Webbink}(1985)}]{Webbink_GC_90}
{Webbink}, R.~F. 1985, in IAU Symp. 113: Dynamics of Star Clusters,  541--577

\bibitem[{{Wilkinson et~al.}(2003)}]{Wilkinson_2003}
{Wilkinson}, M.~I. {et al.} 2003, MNRAS, 343, 1025

\bibitem[{{Woitas} {et~al.}(2001){Woitas}, {Leinert}, \& {K{\"
  o}hler}}]{Woitas_MassRatio_01}
{Woitas}, J., {Leinert}, C., \& {K{\" o}hler}, R. 2001, A\&A, 376, 982

\bibitem[Zoccali et al.(2001)]{2001ApJ...553..733Z} Zoccali, M., et al.\ 
2001, ApJ, 553, 733 

\end{thebibliography}
\end{document}